\documentstyle[psfig,astrobib,subfigure]{mn-ab}
\newcommand{\kms}{\mbox{\,km\,s$^{-1}\,$}}
\newcommand{\Kkms}{\mbox{\,K\,km\,s$^{-1}\,$}}

\newcommand{\Msun}{\,M$_{\odot}~$}
\newcommand{\Lsun}{\,L$_{\odot}$}

\newcommand{\HII}{\mbox{$\mathrm{H\,{\scriptstyle II}}\,$}}

\newcommand{\hh}{$\mathrm{H_{2}}$}
\newcommand{\coh}{\mbox{$^{12}${\rmfamily CO}{(2--1)}}}
\newcommand{\col}{\mbox{$^{12}${\rmfamily CO}{(1--0)}}}
\newcommand{\co}{$^{12}${\rmfamily CO}$\,$}

\newcommand{\be}{\begin{equation}}
\newcommand{\ee}{\end{equation}}
\newcommand{\um}{$\mu$m}
\newcommand{\nCar}{\mbox{$\eta$ Car}}
\newcommand{\touter}{T$_{\mathrm outer}$}
\newcommand{\tinner}{T$_{\mathrm inner}$}
\newcommand{\rmin}{R$_{\mathrm min}$}
\newcommand{\rmax}{R$_{\mathrm max}$}

\newcommand{\MSX}{{\it MSX}}
\newcommand{\IRAS}{{\it IRAS}}
\title[PDRs and star formation in the Carina Nebula]{Photodissociation regions and star formation in the Carina Nebula}
\author[Rathborne,  Burton, Brooks, Cohen, Ashley \& Storey]
       {J.M. Rathborne$^{1}$, M.G. Burton$^{1,2}$, K.J. Brooks$^{3}$, M. Cohen$^{4}$,
\newauthor M.C.B. Ashley$^{1}$, J.W.V. Storey$^{1}$\\
$^{1}$School of Physics, The University of New South Wales, Sydney, NSW,
2052, Australia\\
$^{2}$School of Cosmic Physics, Dublin Institute for Advanced Studies, 5 Merrion Square, Dublin 2, Ireland\\
$^{3}$European Southern Observatory, Casilla 19001, Santiago 19, Chile\\
$^{4}$Radio Astronomy Lab, 601 Campbell Hall, University of California, Berkeley, CA 94720, USA}

\pagerange{\pageref{firstpage}--\pageref{lastpage}}
\pubyear{2002}

\begin{document}

\maketitle
\label{firstpage}
\begin{abstract}
We have obtained wide-field thermal infrared (IR) images of the Carina
Nebula, using the SPIREX/Abu telescope at the South Pole. Emission from
poly-cyclic aromatic hydrocarbons (PAHs) at 3.29\um, a tracer of photodissociation 
regions (PDRs), reveals many interesting well defined clumps and diffuse regions
throughout the complex. Near-IR images (1--2\um), along with images from the
Midcourse Space Experiment (\MSX) satellite (8--21\um) were incorporated to study the 
interactions between the young stars and the surrounding molecular cloud in more
detail. Two new PAH emission clumps have been identified in the Keyhole Nebula 
and were mapped in \coh\, and (1--0) using the SEST. Analysis of their physical 
properties reveals they are dense molecular clumps, externally heated with PDRs on their 
surfaces and supported by external pressure in a similar manner to
the other clumps in the region. A previously identified externally heated 
globule containing IRAS 10430-5931 in the southern molecular cloud, shows strong 3.29-, 8-, 
and 21-\um\,emission, the spectral energy distribution (SED) revealing the location of an 
ultra-compact (UC) \HII\,region. The northern part of the nebula is complicated, with PAH emission 
inter-mixed with mid-IR dust continuum emission. Several point sources are located here and 
through a two-component black-body fit to their SEDs, we have identified 3 possible UC \HII\, 
regions as well as a young star surrounded by a circumstellar disc. This implies that star 
formation in this region is on-going and not halted by the intense radiation from the 
surrounding young massive stars.
\end{abstract}

\begin{keywords}
\HII\, regions -- stars: formation -- ISM: lines and bands -- ISM: molecules -- ISM: structure.
\end{keywords}

\section{Introduction}
The Carina Nebula is an \HII\, region/molecular cloud complex containing some
of the most massive star clusters identified in our galaxy. The nebula is at a distance of 
2.2~kpc \cite{Tovmassian95} and allows us to study the impact young massive stars have on 
their surroundings and in particular whether they trigger or hinder further star formation.
The region is immersed in the UV radiation and stellar winds from the many clusters; 
Trumpler 14 (Tr 14) and Tr 16 being the most influential. Several O-type stars are found 
here, including six O3 stars \cite{Walborn95} and the spectacular star \nCar. Optical images 
reveal the many interesting aspects of this region including bright-rimmed globules, 
filaments, dark patches and large dust lanes that bisect the complex. Radio continuum 
observations reveal a large ionized region with two peaks, Car~I and Car~II \cite{Gardner68}. 
Both of these \HII\, regions contain arc and filamentary structures many of which correspond to 
optical features. They are consistent with ionization fronts from Tr 14 and Tr 16 respectively 
(\citeNP{Retallack83}; \citeNP{Whiteoak94}; \citeNP{Brooks01}). \par
Evidence for on-going star formation in the nebula has been scarce, with only one 
possible site identified \cite{Megeath96}. Due to the paucity of star formation
sites, it was suggested that the massive stars were too destructive on their 
environment, thus hindering any further star formation \cite{Cox952}. 
This view has been recently challenged by results from the Midcourse Space 
Experiment (\MSX), revealing several embedded infrared (IR) sources where 
star formation may be active \cite{Smith00}. In addition, Brooks et al. (2001) have 
identified two compact \HII\, regions possibly corresponding to very young O-type stars.

Recent results from the Short Wavelength Spectrometer (SWS) on-board the Infrared Space 
Observatory (ISO) in the vicinity of the Car~I \HII\, region, reveal a ``broad 22-\um'' 
feature extending from 18--28\um~\cite{Chan00}. This feature has also been reported in the 
M17-\HII\, region \cite{Jones99}, and is similar to an emission band attributed to freshly 
formed dust in ejecta from Cas A \cite{Arendt99}. The spatial variations in intensity 
with respect to the distance from the Car~I emission peak suggests heating of interstellar 
dust. Arendt et al. (1999) propose the broad 22-\um\, feature arise from Mg protosilicate grains.

Two distinct regions of CO emission have been identified in the Carina Nebula 
(\citeNP{deGraauw81}; \citeNP{Whiteoak84}; \citeNP{Brooks98}). Large-scale studies 
have found these are part of a much larger giant molecular cloud (GMC). The two 
regions, referred to as the northern and southern clouds, are 
elongated over 130\,pc and contain a mass in excess of 5$\times$ 10$^{5}$ \Msun 
\cite{Grabelsky88}. Between them lies the Keyhole Nebula, consisting of many discrete molecular 
clumps with masses of {\mbox {$\sim$ 10\Msun}} \cite{Cox951}. Little is known about the extent 
of the interaction between the stars and molecular material. The far-UV (FUV) radiation 
from massive stars penetrates through an inhomogeneous GMC and dominates the heating and 
chemistry of the gas. This occurs at interface regions where the molecular material is 
directly exposed, resulting in photo-dissociation regions (PDRs; \citeNP{Hollenbach99}). Under 
this premise we would expect PDRs to be prominent throughout the Carina Nebula.

Poly-cyclic aromatic hydrocarbon (PAH) molecules are excellent tracers of PDRs. Emission from 
PAH molecules is widespread in many astrophysical environments (\citeNP{Cohen86}; 
\citeNP{Allamandola89}; \citeNP{Cohen89}). PAH molecules are excited by FUV radiation and 
emit fluorescently in several IR bands (often referred to as the Aromatic Infrared Bands, 
AIBs). The main features which occur at 3.3, 6.2, 7.7, 8.6 and 11.3\um\,are thought to arise 
from various bending and stretching modes of PAH molecules (\citeNP{Leger84}; 
\citeNP{Allamandola85}; \citeNP{Geballe94}; \citeNP{Joblin95}). As well as displaying strong 
IR emission features, continuum emission from 1--5\um\, has also been found in association with 
these molecules \cite{Sellgren83}. Recent results from ISO have contributed to advances in
modelling PAH cation emission (see review by \citeNP{Verstraete01}), including the 
discovery of a new feature at 16.4\um\, \cite{Moutou00}.

Our objective in this paper is to use the band emission from PAH molecules in the Carina 
Nebula to study the impact of the FUV radiation from the massive stars on the surrounding 
molecular material. To do this we obtained both high-resolution and wide-field images in the 
thermal IR of a large region across the Carina Nebula. These images were combined with near- 
and mid-IR images and with results from follow-up molecular line observations.

\section{Observations}
\subsection{Thermal Infrared: 3--4\um}
\label{spirex-obs}
Observations were obtained across the Carina Nebula using the South Pole 
Infra-Red EXplorer (SPIREX) during 1999. SPIREX was a 60-cm telescope built and operated at 
the South Pole from 1994 to 1999 \cite{Hereld90}. It was equipped in 1998 with the NOAO Abu 
IR camera \cite{Fowler98}, which incorporated an engineering grade \mbox{1024 $\times$ 1024} 
`Aladdin' InSb 1--5\um\, array. This produced circular images of approximately 10 arcmin 
diameter, with a 0.6 arcsec pixel scale. 

The Carina region was imaged using two filters, a narrow-band PAH filter centred on 3.29\um\,
(half-power width of 0.074\um) and a continuum L-band filter centred on 3.5\um\, 
(half-power width of 0.618\um). Five overlapping positions across the centre of the nebula were
obtained in both these bands. Integration times of 60\,s and 6\,s were used 
for individual frames respectively. The observing sequence in all cases consisted of 
a set of sky frames followed by two sets of object frames. Each set included a 5-point cross, 
with each frame offset slightly from the previous.

All data reduction was achieved using IRAF\footnote{IRAF is the Image Reduction and Analysis Facility written and supported by NOAO, see http://iraf.noao.edu/}, with archived images 
used for dark subtraction and flat-fielding. Sky-subtraction was performed
using the 6 frames nearest in time to the object frame (regardless of the 
actual image type). All frames covering the same region were then
registered and combined to form the final image. Common stars in adjacent
frames were used to match the overlapping regions to produce a larger mosaic.
All registering and combining of the individual frames and of the final 
mosaic were achieved using routines written by \citeN{McGregor95}.
A coordinate axis was added using a corresponding image obtained with the 
Digitised Sky Survey\,\footnote{see http://archive.stsci.edu/dss} (DSS) and 
the program {\sc koords} \cite{Gooch96}.

For the PAH-band a total of 72 individual frames were used to produce the final mosaiced 
image. This image contains both line and continuum emission. The amount of continuum
contribution in the PAH band was estimated using the L-band images and vice-versa for 
contamination to the continuum fluxes from PAH molecules. This contamination was  
removed from the PAH emission and L-band continuum flux values given for each source.

Observations of the standard star, HR\,4638, using both the PAH- and L-band filters, were 
used for flux calibration (B3V star, L-band magnitude of 4.50\,mag). Based on the 
repeatability of the standard over the observing periods, we estimate the PAH fluxes to have 
an uncertainty of \mbox{$\pm$ 9 per cent}, with an uncertainty of $\pm$ 20 per cent 
for the L-band fluxes. The L-band fluxes are more uncertain due to poorer 
weather conditions and tracking during the observations. The diffraction
limit at \mbox{3.3\um\,} of the telescope is 1.4 arcsec, comparable to
the typical ice-level seeing of the site. However, a combination of
tracking errors, tower shake and co-addition of frames limited the stellar
FWHM for the PAH images to 2.8 arcsec. The final PAH image has a 
{\mbox {1 $\sigma$ rms noise of 4$\times$10$^{-15}$ erg s$^{-1}$ cm$^{-2}$ pixel$^{-1}$}}. 

\subsection{Mid Infrared: 8--21\um}
\label{msx-obs}
Degree-scale images from 8--21\um\, were obtained toward the Carina region from the \MSX\, 
satellite\footnote{For a full description of the \MSX\, satellite, the astrophysical experiments and the observing techniques see \citeN{Mill94}, Price et al. (1996), \citeN{Egan98} and \citeN{Price01}.} 
and covered four discrete bandpasses; Band A (6.8--10.8\um, hereafter referred to as the 
8-\um\, band), Band C (11.1--13.2\um), Band D (13.5--15.9\um) and Band E (18.2--25.1\um, 
hereafter referred to as the 21-\um\, band). These images reveal a great deal of information 
about the interaction and physical conditions of the material in this region. Emission seen 
in the 21-\um\, band arises from warm dust at $\sim$ 100\,K, and is used to identify 
deeply embedded sources. The 22-\um\, feature identified by \citeN{Chan00} contributes 
$\sim$15--30 per cent of the flux in the 21-\um\, band. This was estimated using their Figs.~1 
and 2, in which the spectra represent the position where the feature appears the strongest.

The 8-\um\, band is more complex, including several discrete PAH emission features, the 
plateau of emission from 6--9\um\, as well as emission from dust at $\sim$ 400\,K. It is not 
possible to determine which process causes the observed emission from this band alone. By 
using it in combination with the 3.29- and 21-\um\, emission structures, it is possible to 
identify regions where the 8-\um\, emission is coming from PAH molecules and where it arises 
from warm dust. In general, where the 3.29- and 8-\um\, emission are spatially well-correlated 
we attribute the emission dominantly to that from PAHs. If the 8- and 21-\um\, emission are 
spatially well-matched, however, the emission is dominated by continuum emission.  

\subsection{Near Infrared: 1--2\um}
Bright extended emission detected in the SPIREX/Abu images toward the southern Carina
region and several point sources located further north  were observed using the 
Australian National University 2.3-m telescope at Siding Spring 
Observatory, Australia. The Near-IR camera CASPIR (Cryogenic Array Spectrometer Imager; 
\citeNP{McGregor94}) contains a 256 $\times$ 256 InSb array and was used
to image \mbox{J (1.25\um)}, H (1.65\um) and K (2.2\um) bands during 2000 June.
Individual images cover approximately {\mbox {2 $\times$ 2 arcmin$^{\,2}$}} (0.5 arcsec 
pixel scale) with integration times of 5\,s for J- and H-band and 2\,s for K-band images.
Standard stars were observed before each set of image frames.

Data reduction was achieved using CASPIR specific routines within IRAF \cite{McGregor95}. 
Images were dark- and bias-subtracted, before linearisation and flat-fielding
were performed. Sky background was removed by averaging the sky value obtained over several 
consecutive image frames. Coordinates were applied using a corresponding image from the
DSS and the program {\sc koords}.

\subsection{Millimetre}
Two new PAH emission structures located in the Keyhole region were identified from 
the images obtained with SPIREX/Abu. Observations of \col\, at 115.271\,GHz and 
\coh\, at 230.537\,GHz were obtained toward these structures during 
2000 October using the 15-m Swedish-ESO Submillimetre Telescope 
(SEST)\footnote{The SEST is operated jointly by ESO and the Swedish National Facility for Radio Astronomy, Onsala Space Observatory, Chalmers University of Technology.} at La 
Silla Observatory, Chile. Both transitions were observed simultaneously
with the IRAM-built 230/115\,GHz SIS receiver tuned to single side-band 
(SSB) mode and connected to a narrow-band (43\,MHz) 
Acousto-Optical Spectrometer (AOS) with a channel resolution of 0.08\,MHz. This 
corresponds to approximately 0.1\kms at 115\,GHz and 0.05\kms at 230\,GHz. The 
bands were centred on a radial velocity of $-$10\kms with respect to the local standard of 
rest (LSR) for one emission structure and $-$5\kms for the other. Position-switching mode was 
used with \nCar\, used as the signal-free reference position. An integration time of 
60\,s was used for all observations. The average system temperatures during the 
observations were $\sim$ 200 K for the 115\,GHz band and $\sim$ 400 K for the 
230\,GHz band. Chopper-wheel calibration was performed every 10\,mins
to obtain atmosphere-corrected antenna temperatures according to the method described by 
\citeN{Ulich76}. The telescope pointing and sub-reflector focusing were checked regularly 
using suitably bright nearby SiO masers. We estimate the pointing accuracy to be better than 
10 arcsec and adopt the standard SEST value of 10 per cent uncertainty in the temperature
scale. The beam size of the SEST is 45 arcsec at 115\,GHz and 22 arcsec at 230\,GHz.

Observations toward both PAH structures were obtained using a 20 arcsec pointing grid,
covering areas of $\sim$~4~$\times$~4~arcmin$^2$ and $\sim$~3~$\times$~3~arcmin$^2$ for
each region. These regions were centred on the peak in the PAH emission.
%(see Table~\ref{pah-features}).

All spectra were processed using {\sc gildas} software \cite{Buisson99}. Linear 
baselines were initially removed from each spectra and the temperature scale 
converted to main-beam brightness temperature using the values for the main 
beam efficiencies of 0.7 for 115\,GHz and 0.5 for 230\,GHz. 
The \col\, data were smoothed to a {\mbox{40 arcsec}} grid to match the beam size of the
SEST at 115\,GHz. The average rms noise per spectral channel was 0.8 K for both
\col\, and \coh.

\section{Results and Discussion}

\subsection{The Carina Nebula in 3.29-\um~emission}

\begin{figure*}
\psfig{file=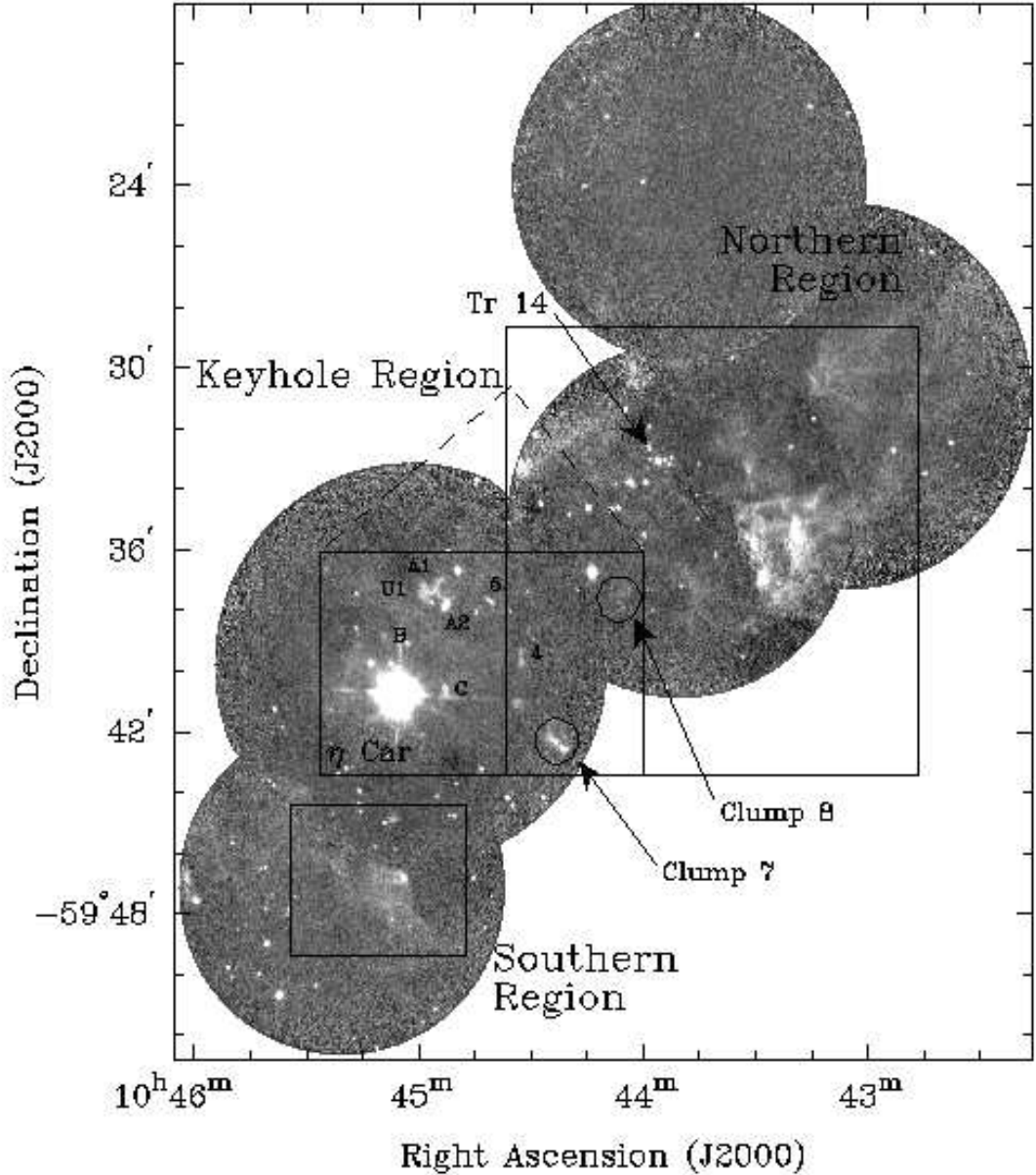,height=16.5cm}
\caption{\label{pah-mosaic}The Carina Region in the 3.29-\um\, PAH and continuum emission. The cluster Tr 14 is apparent along with the massive star \nCar. Clumps of 
emission within the Keyhole region are labelled along with the regions (outlined by boxes) discussed in more detail in the following sections.}
\end{figure*}

\begin{table*}
\begin{centering}
\caption{\label{pah-features}Measured 3.29-\um\, PAH emission parameters for the regions shown in Fig.~\ref{pah-mosaic}. These represent continuum corrected fluxes$^{a}$.}
\begin{minipage}{130mm}
\begin{centering}
\begin{tabular}{@{}lccccc}
\hline
Region	& \multicolumn{2}{c}{Peak position} & Peak Intensity$^{b}$ & Size 
	& Integrated Intensity \\
& RA   & DEC & ($\times$ 10$^{-14}$ erg s$^{-1}$ & \arcsec\ $\times$ \arcsec
	&($\times$ 10$^{-11}$ erg \\
& (2000) & (2000) &cm$^{-2}$ arcsec$^{-2}$) &
	&s$^{-1}$ cm$^{-2}$) \\
\hline
Clump 7 & 10 44 23 & $-$59 42 14 &  7.1 	&  90  $\times$ 60   &  4.0 \\
Clump 8 & 10 44 06 & $-$59 37 21 &  3.4 	&  60 $\times$ 100   &  2.9 \\
Southern region    & 10 45 00 & $-$59 47 03 &  5.1 	&  300 $\times$ 120  &  3.1 \\
Northern region    & 10 43 20 & $-$59 34 53 &  35 	&  450 $\times$ 240  &  97 \\
\hline
\end{tabular}\\
\end{centering}

$^{a}$~The contamination to the total fluxes from continuum emission was estimated to be 10~per~cent for clump~7 and the Northern region, 20~per~cent for the Southern region and less than 1~per~cent for clump~8.\\
$^{b}$~1 $\sigma$ rms noise of 1$\times$10$^{-14}$ erg s$^{-1}$ cm$^{-2}$ arcsec$^{-2}$\\
\end{minipage}
\end{centering}
\end{table*}

Fig.~\ref{pah-mosaic} shows the 3.29-\um\, emission observed across the
Carina Nebula. The emission can be grouped into three main regions identified as: Keyhole 
region, southern region and northern region. The peak position, peak intensity, size and 
total integrated intensity of the structures within these three regions are given in 
Table~\ref{pah-features}. The contamination from continuum emission was estimated 
and subtracted from the total flux seen toward each region. Thus, the flux values given in 
Table~\ref{pah-features} are estimates of the contribution to the emission from the 3.29-\um\, 
PAH line only. The extended emission seen throughout Fig.~\ref{pah-mosaic} is predominantly from 
the 3.29-\um\, PAH emission line, with the majority of the continuum emission contained in the
point sources.

The region situated at the centre of Fig.~\ref{pah-mosaic} includes
the Keyhole Nebula and \nCar. Previous studies of these sources have shown they are dense 
molecular clumps \cite{Cox951} which, in addition to bright 3.29-\um\, PAH emission, exhibit 
strong \hh\, (2.12-\um) emission \cite{Brooks00}. The coincidence between these features is 
indicative of the existence of PDRs on the clumps' surfaces. The larger area covered in our 
study reveals two additional 3.29-\um\, emission structures. One is located in the south-western 
corner of the Keyhole region, while the other is much fainter and is displaced 
further toward Tr 14. This region will be discussed in more detail in section~\ref{keyhole}.

In contrast to the Keyhole region, the 3.29-\um\, emission detected in the southern region is
much more diffuse. The extent of the emission here traces the sharp northern rim of 
the southern molecular cloud. Several \IRAS\, sources are located along the 
edge of this molecular cloud. The peak seen in 3.29-\um\, emission corresponds to one 
such source, IRAS 10430-5931, and is the location of a bright-rimmed globule 
identified through a combination of molecular line and IR observations \cite{Megeath96}. 
The structures seen in this region will be discussed further in section~\ref{southern-region}.

\begin{figure}
\centering
\psfig{file=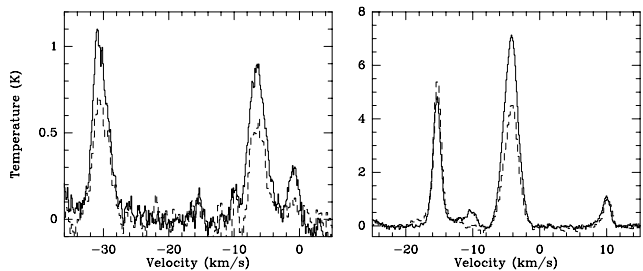,width=0.48\textwidth,angle=-90}
\caption{\label{integrated-intensity}The averaged line profiles for all positions observed toward clumps 7 (left) and 8 (right). The solid line refers to the \coh, while the dashed line represents the (1--0) emission.} 
\end{figure}

The northern region reveals a similar environment to that seen in the south, with
extended diffuse emission covering the whole region. The brightest 3.29-\um\, emission 
is located to the west of the cluster Tr 14 and follows the sharp edge of the northern 
molecular cloud. This edge corresponds to a dark obscuring region in optical images. The 
second peak in the 3.29-\um\, emission is located further to the west. The extent of the 
emission in this region is coincident with the edge of the ionization peak Car~I, confirming 
the existence of a PDR at the interface between the ionized and molecular material. The peak 
intensity of the emission here is almost 10 times higher than the other 
regions observed (see Table~\ref{pah-features}). The close proximity of the cluster to the 
molecular material offers an ideal location in which to study their interaction. 
The northern molecular cloud extends much further to the north of this region than is 
indicated by the 3.29-\um\, emission. The absence of any emission to the north may
be a consequence of a weaker FUV radiation field. The northern region
will be discussed in more detail in section~\ref{northern-region}.

From the 3.29-\um\, emission seen in Fig.~\ref{pah-mosaic} it is clear that PDRs extend
across the whole Carina GMC. The emission we see is not limited to the vicinity of a
single source, but penetrates through the whole region and interacts with the molecular
material across the complex. This is facilitated by the inhomogeneous nature of the region
and is a common characteristic of star forming regions, including NGC 6334 \cite{Burton00}, the
Rosette \cite{Schneider98}, M17 (\citeNP{Stutzki88}; \citeNP{Meixner92}), 
Orion \cite{Tauber95} and W51 \cite{Genzel88}.

\subsection{Keyhole region}
\label{keyhole}

The molecular clumps of this region are typically 10-30 arcsec in length
and all correspond to striking optical features such as dark patches, some
with bright rims. Their mass is of order 10\Msun, they are gravitationally bound
and emit over a velocity from 0 to -30\kms (LSR) \cite{Cox951}.

\subsubsection{New molecular clumps}
\begin{table}
%\begin{minipage}{0.4\textwidth}
\centering
\caption{\label{properties}Properties of the \coh\, and (1--0) profiles seen toward clumps 7 
and 8. These values are obtained from Gaussian fits to the profiles shown in 
Fig.~\ref{integrated-intensity}, where $\Delta$V represents the FWHM estimate for the profiles.}
\begin{tabular}{lccccccc}
\hline
\multicolumn{3}{c}{\col}       &\multicolumn{3}{c}{\coh}    \\
 Velocity & Peak & $\Delta$V  & Velocity & Peak  & $\Delta$V\\
 \kms     & K    &\kms         & \kms     & K     & \kms  \\
\hline	                      			      
\multicolumn{2}{l}{Clump 7}\\
 $-$30.4    & 0.7 & 2.8	& $-$30.5    & 1.0  & 2.8\\        
 $-$15.8    & 0.1 & 1.4	& $-$15.4    & 0.2  & 0.9\\    
 $-$9.8     & 0.2 & 0.5	& $-$10.0    & 0.2  & 0.9\\  
 $-$6.3     & 0.6 & 2.5	& $-$6.5     & 0.9  & 3.0\\      
 $-$1.0     & 0.1 & 1.0	& $-$0.9     & 0.3  & 1.6\\     
\multicolumn{2}{l}{Clump 8}\\
 $-$15.3    & 5.2 & 1.4	& $-$15.3    & 4.7  & 1.4\\      
            &     &    	& $-$10.4    & 0.7  & 1.5\\      
 $-$4.2     & 4.4 & 2.4	& $-$4.3     & 7.0  & 2.5\\      
  ~~10.1    & 1.0 & 1.4 & 9.9        & 1.1  & 1.4\\      
\hline
\end{tabular}
%\end{minipage}
\end{table}

\begin{figure*}
\centering
\mbox{\subfigure[\coh\, emission seen toward clump 7 integrated over the velocity ranges $-$34 to $-$26\kms, $-$9 to $-$2\kms and $-$2 to 2\kms. Contour levels are 4, 8, 12, 16, 20, 24\Kkms.]{\psfig{file=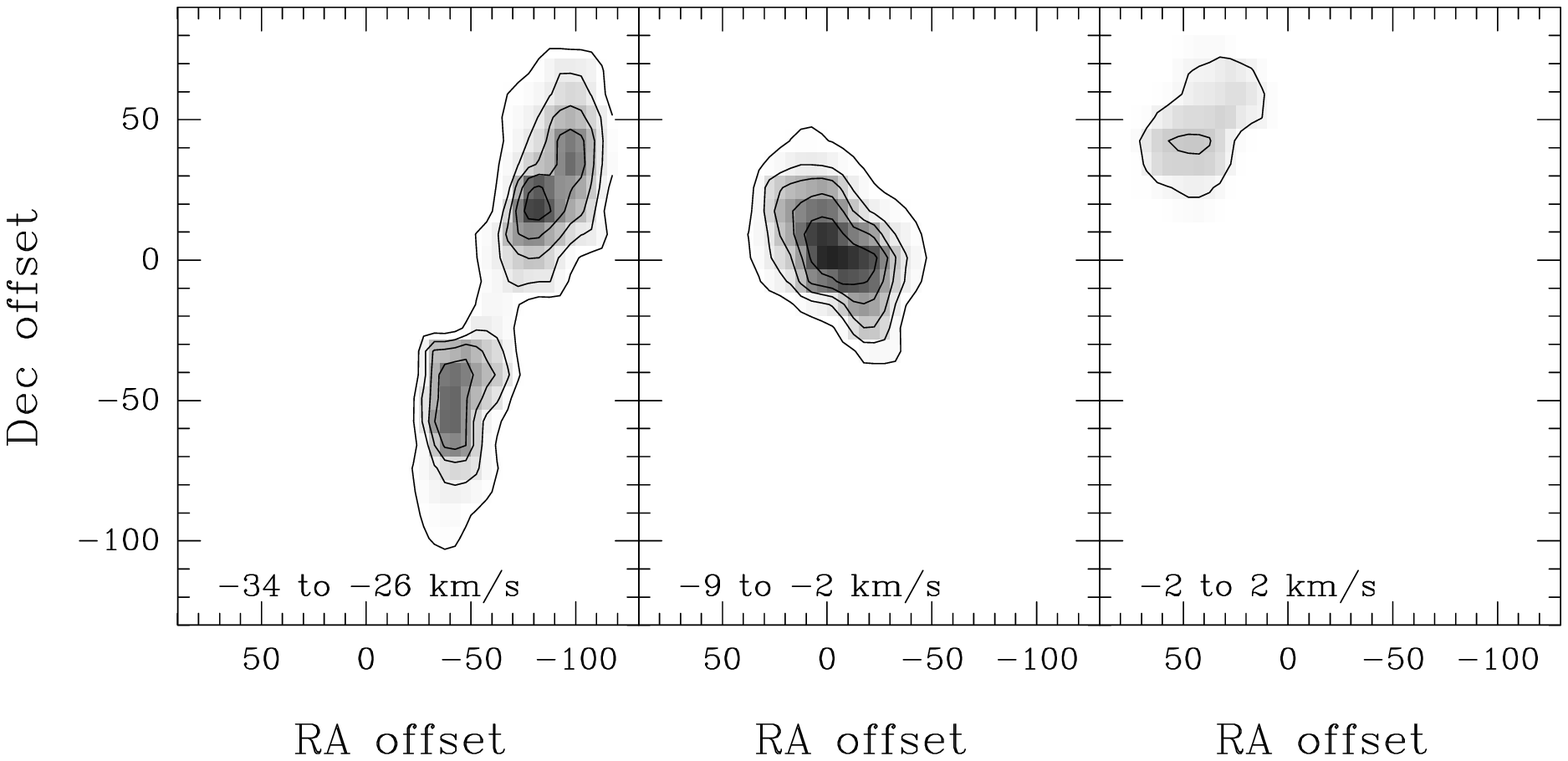,width=0.5\textwidth}}
\subfigure[\coh\, emission seen toward clump 8 integrated over the velocity ranges $-$18 to $-$13\kms, $-$8 to 0\kms and 7 to 13\kms. Contour levels are 12, 14, 16, 18\Kkms; 14, 24, 34, 44, 54, 64, 74\Kkms; and 2, 4, 6\Kkms.]{\psfig{file=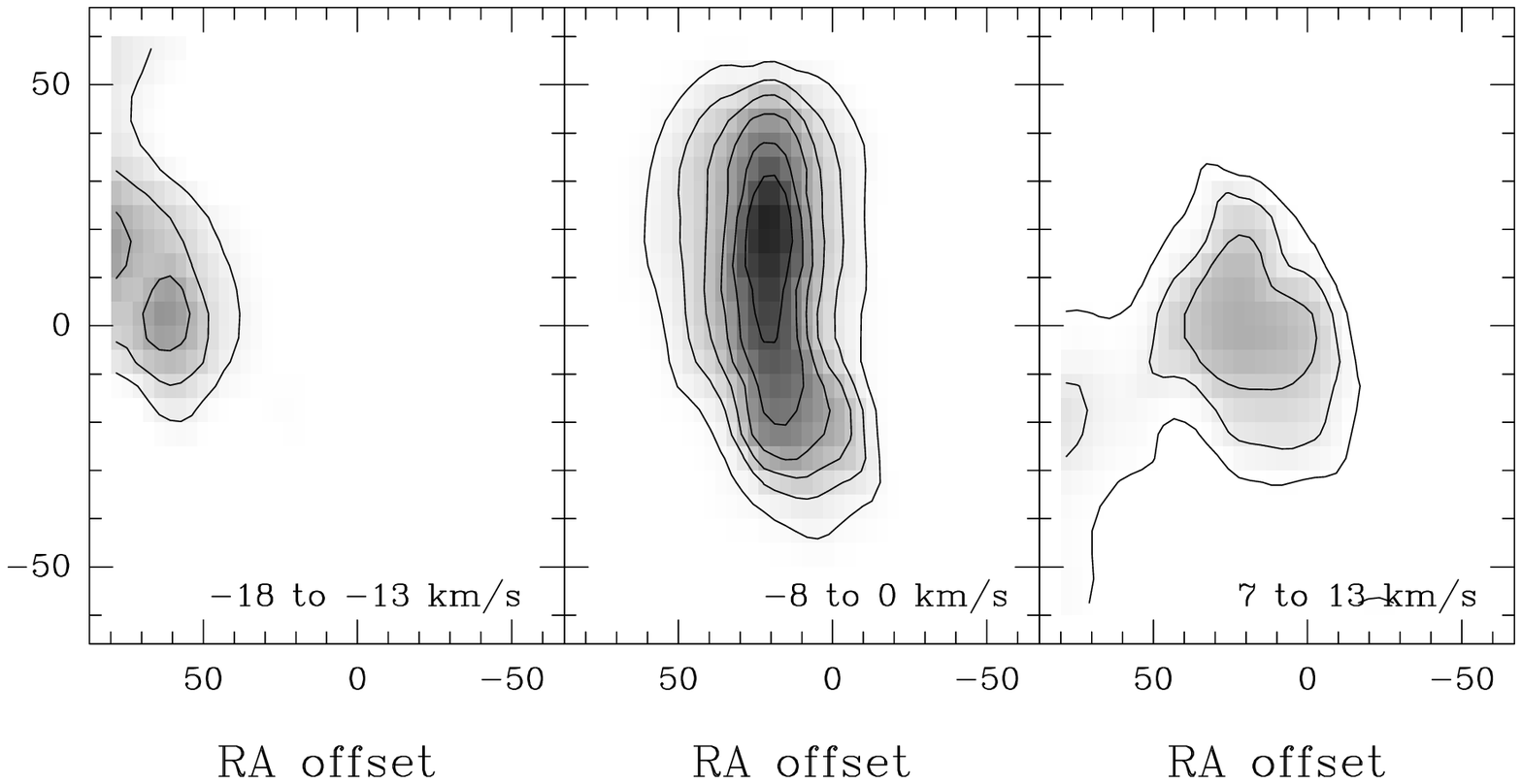,width=0.49\textwidth}}}
\caption{\label{co21}Integrated intensity of the \coh\, emission seen toward the two new 3.29-\um\, PAH emission structures identified in the Keyhole region. The offsets are relative to the 3.29-\um\, emission peaks given in Table~\ref{pah-features}.}
\end{figure*}

\begin{figure*}
\centering
\mbox{\subfigure[3.29-\um\, emission from clump 7 overlaid with \coh\, emission from the component seen in the range $-$9 to $-$2\kms. Contour levels are 4, 8, 12, 16, 20, 24\Kkms.]{\psfig{file=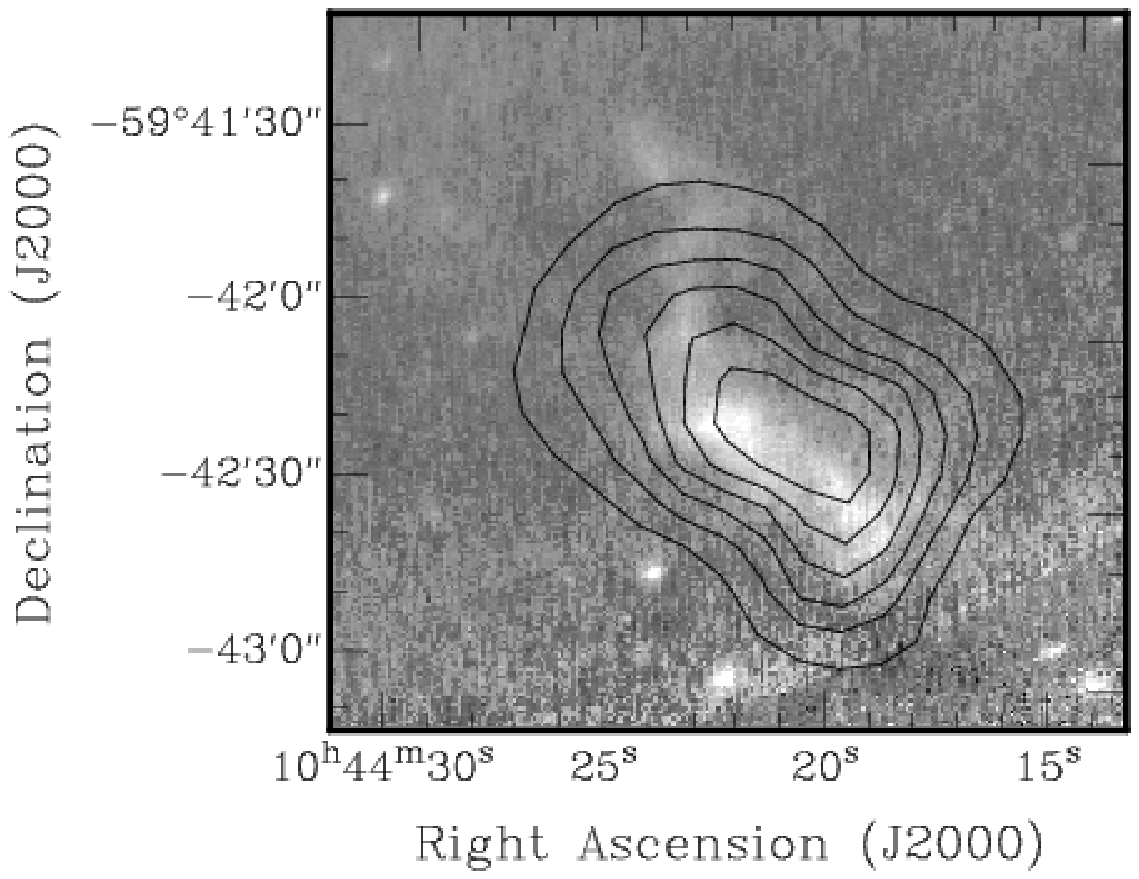,height=7cm,width=0.49\textwidth}}
\subfigure[3.29-\um\, emission from clump 8 overlaid with \coh\, emission for the component seen in the range $-$8 to 0\kms. Contour levels are 14, 24, 34, 44, 54, 64, 74\Kkms.]{\psfig{file=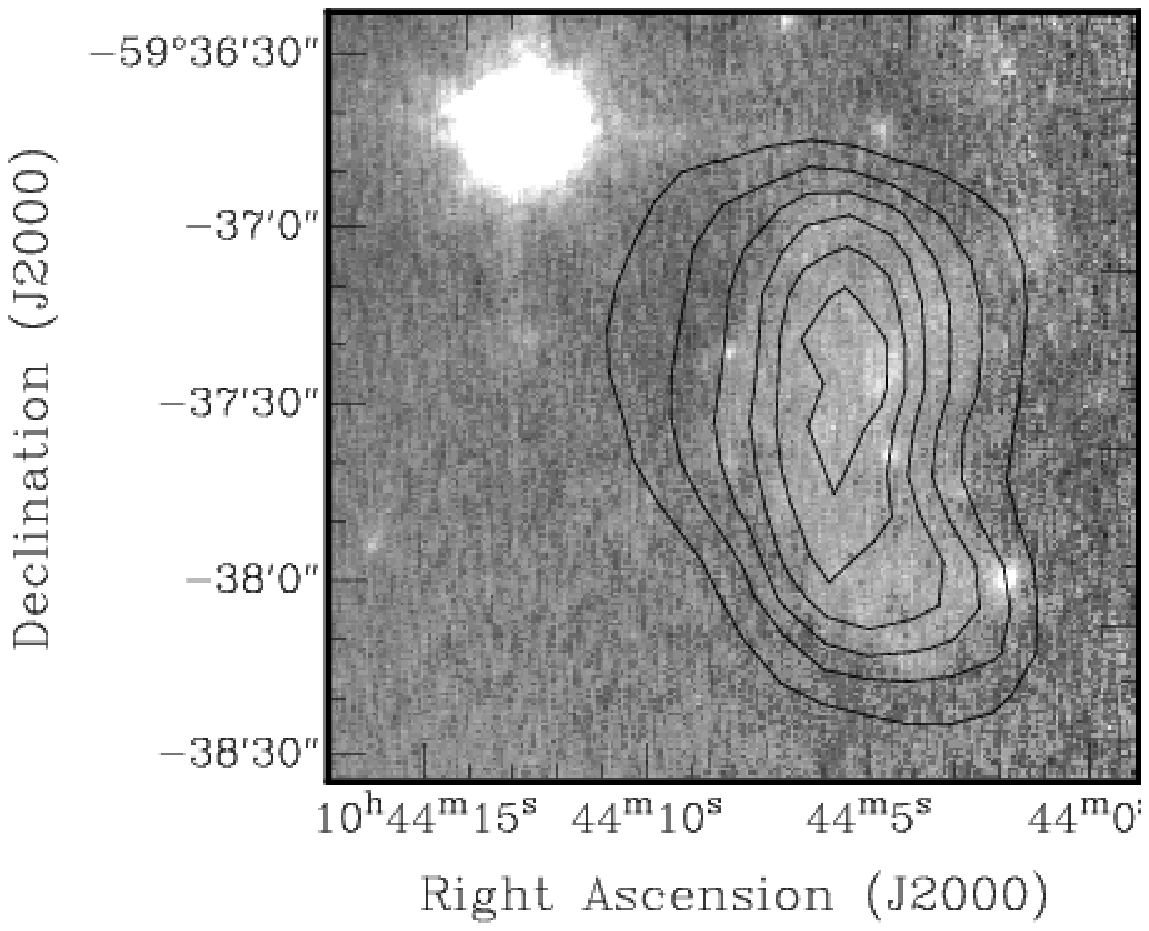,width=0.48\textwidth}}}
\caption{\label{co21-pah}3.29-\um\, emission from the two new clumps identified, overlaid with the corresponding \coh\, emission. Several stars are also evident in these images, in particular at the north east edge in (b).}
\end{figure*}

\begin{table*}
\caption{\label{ratio-mass}The physical properties of the clumps 7 and 8 in the Keyhole region, estimated from values obtained from Gaussian fits to the averaged spectra over the regions enclosed by the clumps, as seen in Fig.~\ref{clump-ratios}. The LTE mass is estimated from local thermal equilibrium conditions. The ratio calculated is for the \co integrated fluxes, with n(\hh) representing the average \hh\, density and  $\Delta$V representing the FWHM of the profile.}
\begin{tabular}{@{}lccccccccccc}
\hline
Clump &\multicolumn{3}{c}{\col}     & \multicolumn{3}{c}{\coh}      & Ratio   & Area      & \multicolumn{2}{c}{Mass} & n(H$_{2}$) \\
      & Velocity & Peak & $\Delta$V & Velocity & Peak & $\Delta$V  & 2--1/1--0 &           & LTE & Virial             &     \\
      & \kms     & K    & \kms      & \kms     & K    &\kms	    &         & deg$^{2}$ & \Msun    & \Msun    & cm$^{-3}$ \\
\hline
7 &$-$6.6 & 1.0 & 2.9 &$-$6.6 & 2.1 & 3.2 & 2.4 & 4.2 $\times$ 10$^{-4}$ & 12 & 417 & 0.7$\times$10$^{3}$\\
8 &$-$4.0 & 7.0 & 2.8 &$-$4.2 & 10.4& 2.1 & 1.1 & 4.6 $\times$ 10$^{-4}$ & 91 & 411 & 4.5$\times$10$^{3}$\\
\hline
\end{tabular}
\end{table*}

In this section we consider the results of the molecular line maps around the
two new 3.29-\um\, PAH emission structures, in order to determine if these new sources are 
in fact molecular clumps with PDRs on their surfaces.

Fig.~\ref{integrated-intensity} shows the \coh\, and (1$-$0) line profiles, 
averaged over the two regions (hereafter referred to 
as clumps 7 and 8). It is clear from the spectra that emission from separate velocity ranges 
exists toward both of these regions. The peak temperature, velocity and FWHM were determined 
from Gaussian fits to each feature and are listed in Table~\ref{properties}. 

From the emission seen toward clump 7, five \coh\, profiles can be identified of
which the strongest and most distinct occur over the ranges $-$34 to $-$26\kms, 
$-$9 to $-$2\kms and $-$2 to 2\kms. Maps of the integrated \coh\, emission over each
of these velocity ranges are shown in Fig.~\ref{co21}(a). The velocity range $-$30 to 
$-$20\kms corresponds to that of the southern molecular cloud (Brooks et al. 1998). The 
emission in the range $-$34 to $-$26\kms is spatially coincident with the edge of this cloud. 
The emission between $-$9 to $-$2\kms  is located further eastwards and corresponds to a 
faint nebulous patch in optical images. The location and morphology of this feature matches 
with the 3.29-\um\, PAH emission (Fig.~\ref{co21-pah}(a)). The final component, in the range
$-$2 to 2\kms, has much fainter molecular emission, with no obvious counterpart in optical 
images of the region.

For the emission seen toward clump 8, four profiles are evident, the strongest of 
these occurring over the ranges $-$18 to $-$13\kms, $-$8 to 0\kms and 7 to 13\kms. 
Fig.~\ref{co21}(b) shows the integrated emission over the velocity ranges seen 
toward this region. The velocity range $-$20 to 0\kms corresponds to the back 
face of the northern molecular cloud \cite{Brooks-phd}. The emission between $-$18 to 
$-$13\kms corresponds the edge of a larger component that extends further northwards and 
eastwards. The emission we see here coincides with bright diffuse emission in optical images 
and is most likely from the \HII\, region in the northern molecular cloud. The emission 
between $-$8 to 0\kms traces an isolated feature that coincides with the detected 3.29-\um\,PAH 
emission (see Fig.~\ref{co21-pah}(b)) and also to a dark optical region. The feature 
seen in the velocity range 7 to 13\kms extends further to the east, with considerably fainter 
emission and no obvious optical counterpart.

Therefore, the two new 3.29-\um\, PAH structures (7 and 8) are consistent with dense molecular 
clumps at velocities of $-$6.3\kms and $-$4.2\kms. The other {\rmfamily CO} emission features 
seen here are most likely foreground and background objects seen along the line of sight, the 
only common features being those at $-$10\kms and $-$15\kms, suggesting these may be much 
larger structures not associated with the Carina Nebula.

\subsubsection{Physical properties of the new clumps}

\begin{figure}
\centering
\psfig{file=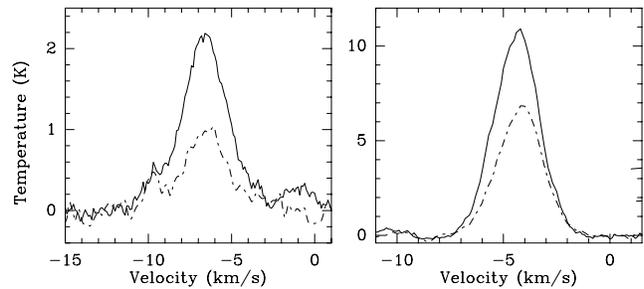,angle=-90,width=0.48\textwidth}
\caption{\label{clump-ratios}The averaged spectra for the regions enclosed by each of the structures; clump 7 (left) and clump 8 (right). The solid line represents \coh\, emission, while the dashed line represents \col\, emission. Results of Gaussian fits to these profiles are given in Table~\ref{ratio-mass}.}
\end{figure}

This section considers the physical conditions within the Keyhole region and the
subsequent exposure of the molecular material. The characteristics of the emission seen
will determine if the new clumps identified here are influenced in a similar
way to the other clumps in the region. To achieve this the spectral profiles contained within 
the region enclosed by the clumps were averaged (Fig.~\ref{clump-ratios}). The results of 
Gaussian fits to these profiles are given in Table~\ref{ratio-mass}, along with the \coh\, 
and (1--0) integrated line ratio, mass estimates and average \hh\, densities.

For clump 7 the \coh\, and (1--0) line ratio was found to be $\sim$ 2, while for clump 8 it was
closer to 1. Ratios greater than unity are typically found in regions were the cloud is warm 
(T$_{k}$~$\geq$~40~K), with average \hh\, densities n(H$_{2}$)~$>$~1~$\times~10^{3}$~cm$^{-3}$ 
and optically thick ($\tau_{21}$~$>$~5) \cite{Sakamoto94}. This ratio is seen in many 
star forming regions including Orion, the Rosette complex and M17 (\citeNP{Castets90}; 
\citeNP{Sakamoto94}; \citeNP{Hasegawa96}) and is explained in terms of the external heating 
of molecular material from FUV radiation. 

Estimates of the local thermal equilibrium (LTE) and Virial masses were obtained for each
of the clumps. The LTE mass was calculated under the assumption that the \hh\, column density
(N(\hh)) is proportional to the integrated intensity of the \col\, line. Using the constant of 
proportionality of {\mbox {3 $\times$ 10$^{20}$(\Kkms)$^{-1}$}} \cite{Scoville91} and assuming 
a distance to the clumps of 2.2~kpc, we estimate values of 12\Msun and 91\Msun for the clumps 
7 and 8 respectively. Calculations for the Virial mass assume that the velocity width of
the {\rmfamily CO} line is a measure of the motion of the overall gas and therefore the mass.
Values obtained for the Virial mass are significantly greater than those calculated 
for the LTE mass, even considering that the LTE values may be under-estimated by a factor of 
3. This implies that the clumps are probably not gravitationally bound and instead are 
supported by external pressure. This is consistent with the other clumps in the region.

\citeN{Brooks00} have developed a geometrical model for the Keyhole region whereby
clumps with velocities of $-$28\kms and more negative are dark obscuring regions in 
front of the nebula; clumps with velocities in the range $-$25 to $-$17\kms are inside the 
nebula and have bright rims in optical images; and clumps with velocities of $-$9\kms and 
more positive are behind the nebula and correspond to faint optical patches. From our results 
we see that the two new clumps with velocities of $-$6\kms and $-$4\kms both correspond 
to dark regions in optical images. According to the above model, this places them at the back
of the nebula.

Estimations of the winds and radiation from the massive stars in particular \nCar\, suggest 
the clumps are photo-evaporating and represent the swept up remains of the natal cloud from 
which the stars formed (\citeNP{Cox951}; \citeNP{Brooks00}). The two new clumps identified 
here support this view although their larger size and higher PAH fluxes suggest the 
photo-evaporation process may be slower. This is consistent with \nCar\, being the most
influential source, their larger displacement from it most likely reducing the severity of the 
interaction.

\begin{figure*}
\centering
\mbox{\subfigure[Clump A1 and A2. Black; 2.5, 2.9, 3.3, 3.7, 4.3 and White; 1.9, 2.0 $\times$ 10$^{-5}$ Wm$^{-2}$\,sr$^{-1}$.]{\psfig{file=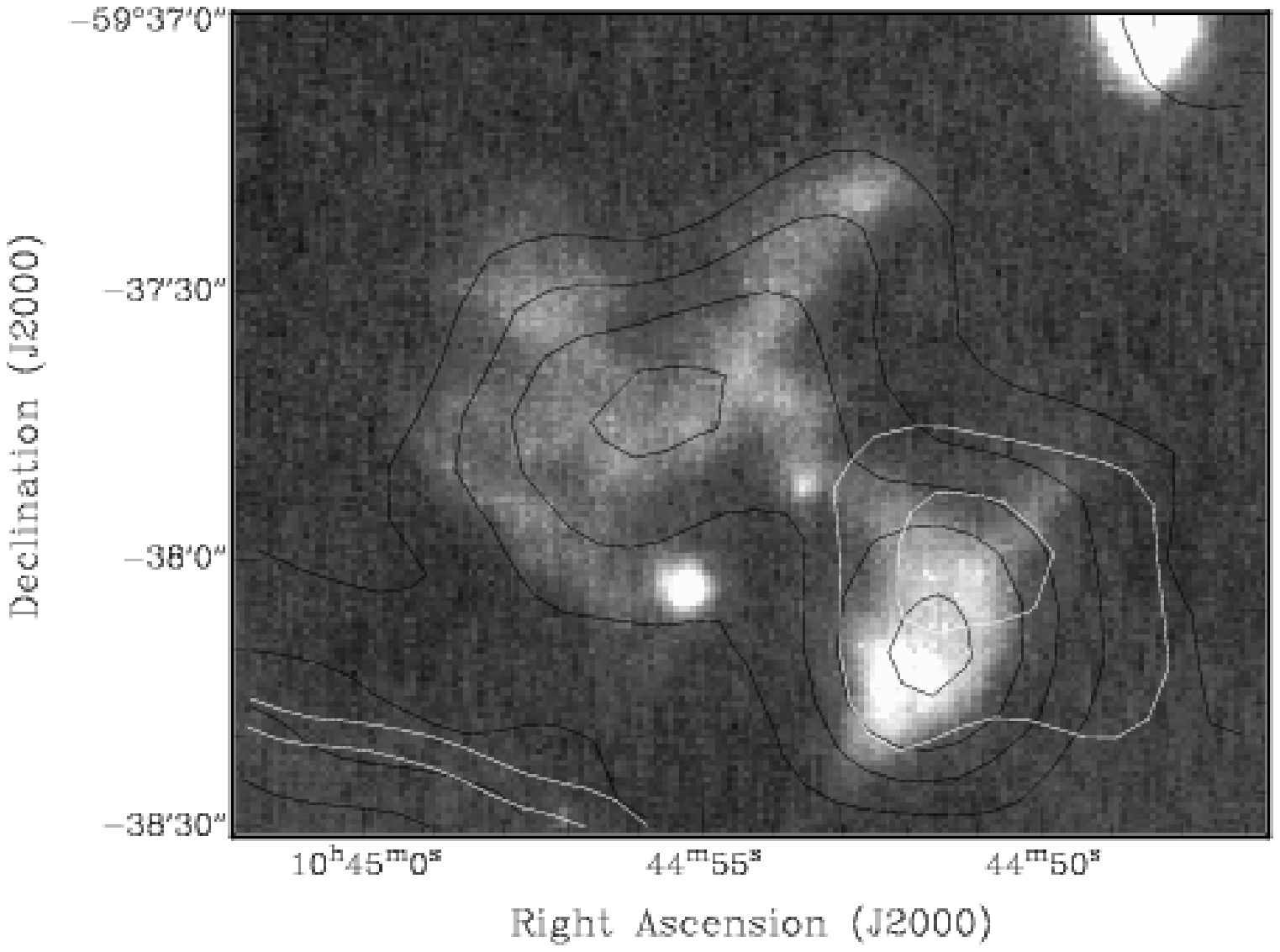,height=5.0cm,width=0.32\textwidth}}
\subfigure[Clump 5. 2.5, 2.7, 2.9, 3.0, 3.1 $\times$ 10$^{-5}$ Wm$^{-2}$\,sr$^{-1}$.]{\psfig{file=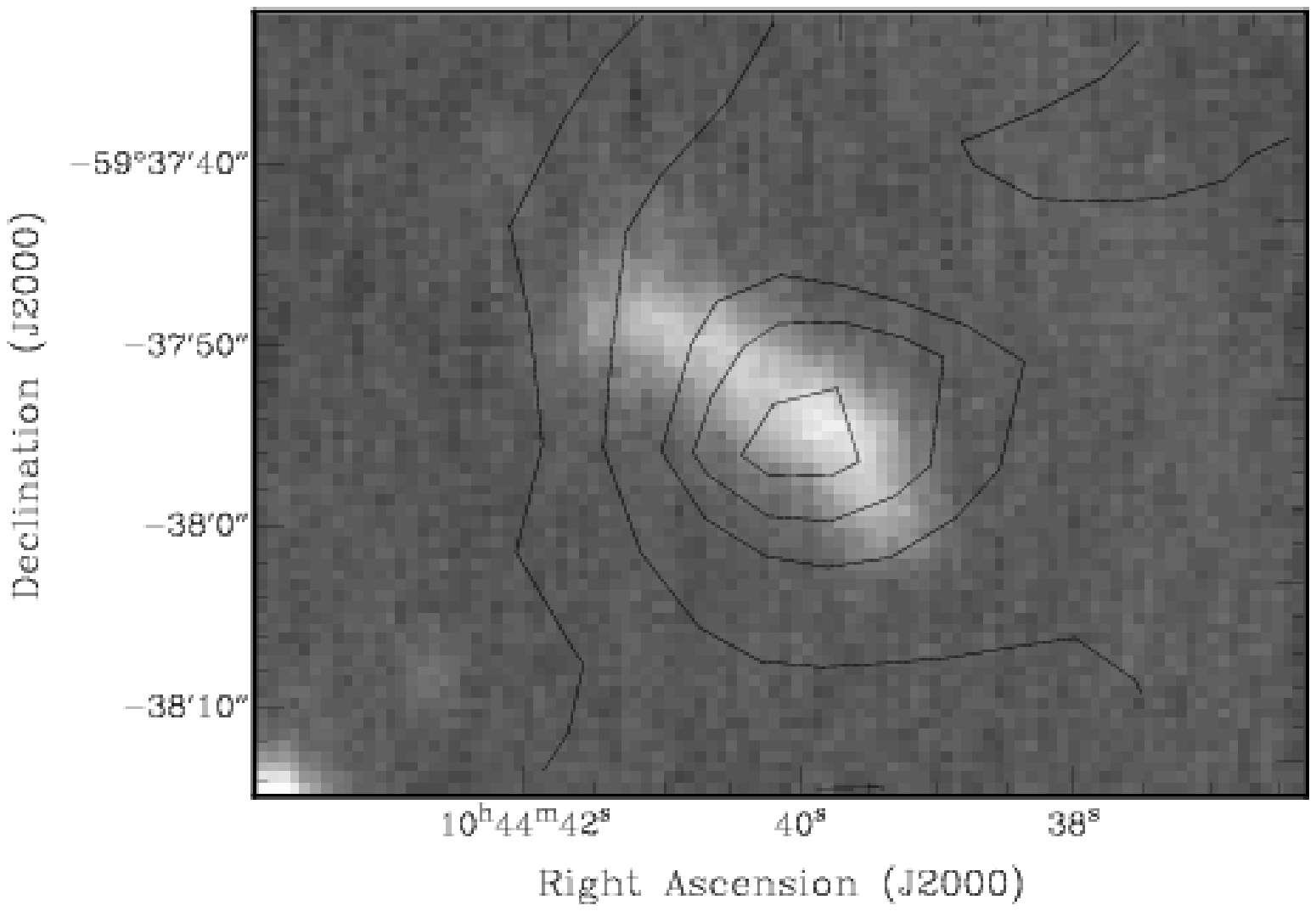,height=5cm,width=0.31\textwidth}}
\subfigure[Clump 7. 2.3, 2.6, 2.9, 3.4 $\times$ 10$^{-5}$ Wm$^{-2}$\,sr$^{-1}$.]{\psfig{file=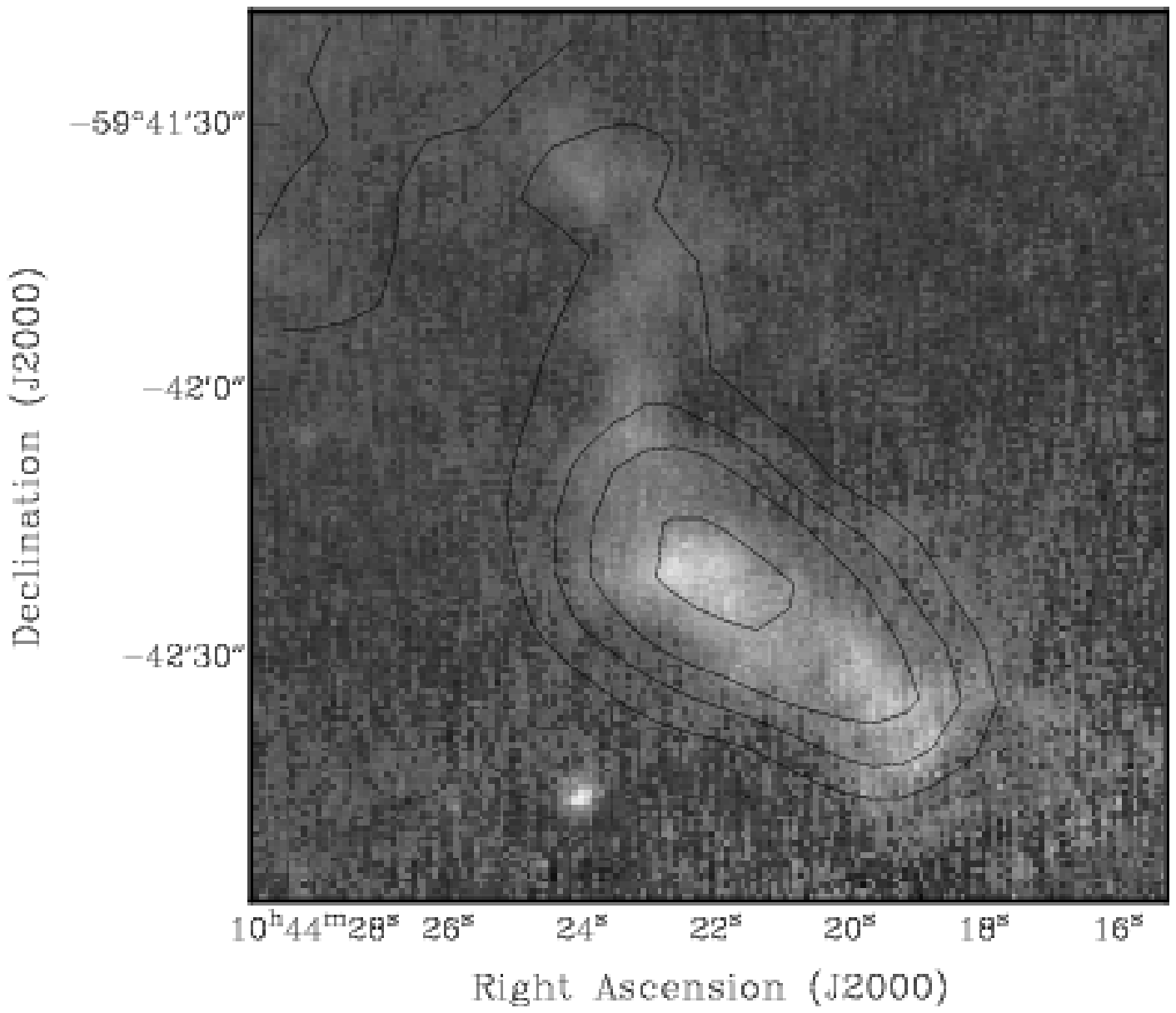,height=5cm,width=0.31\textwidth}}}
\caption{\label{keyhole-pah-A}3.29-\um\, emission in grey scale overlaid with 8-\um\, emission (black contours) seen toward several of the molecular clumps identified in the Keyhole region. Clump A2 also shows 21-\um\, emission (white contours) coincident with it.}
\end{figure*}

\subsubsection{Mid-IR emission in the Keyhole region}

Due to the resolution of the \MSX\, images and the fact that \nCar\, is an extremely strong 
mid-IR source, a study of the Keyhole region is difficult using these data. 
Fig.~\ref{keyhole-pah-A} shows 3.29- and 8-\um\, emission seen toward clumps A1, A2, 5 and 
7. All of these clumps are located far enough from \nCar\, to
be detected above scattered emission from \nCar. Clump 8 is not included however as the 
mid-IR emission is too faint to be distinguished from the many other sources.

\begin{figure*}
\centering
\mbox{\subfigure[3.29-\um\, emission in grey scale with contours of 8-\um\, emission. Levels are 1.2, 1.5, 2, 2.5, 3, 3.5 $\times$ 10$^{-5}$\,Wm$^{-2}$\,sr$^{-1}$.]{\psfig{file=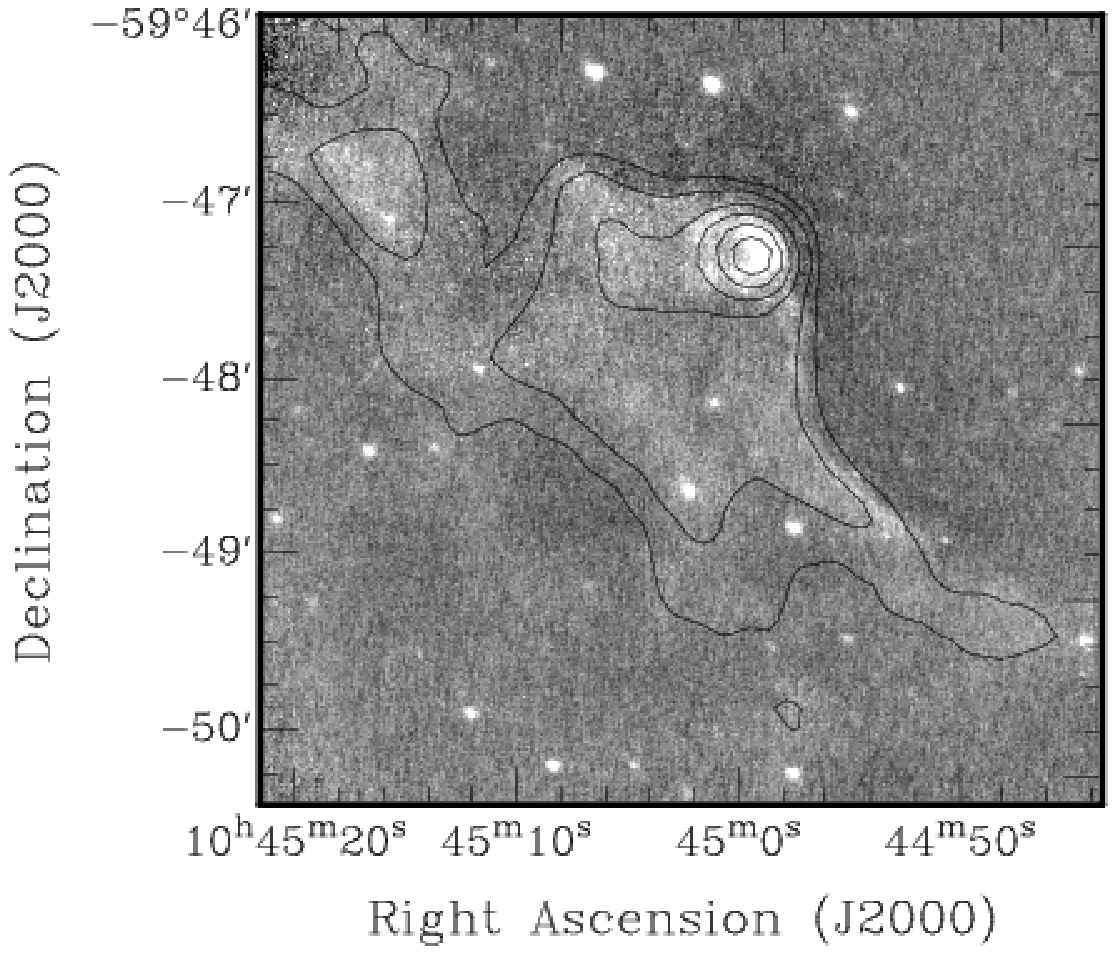,width=0.52\textwidth}}
\hspace{-0.3cm}
\subfigure[8-\um\, emission in grey scale with contours of 21-\um\, emission. Levels are 1.5, 2, 3, 4, 5, 6 $\times$ 10$^{-5}$\,Wm$^{-2}$\,sr$^{-1}$.]{\psfig{file=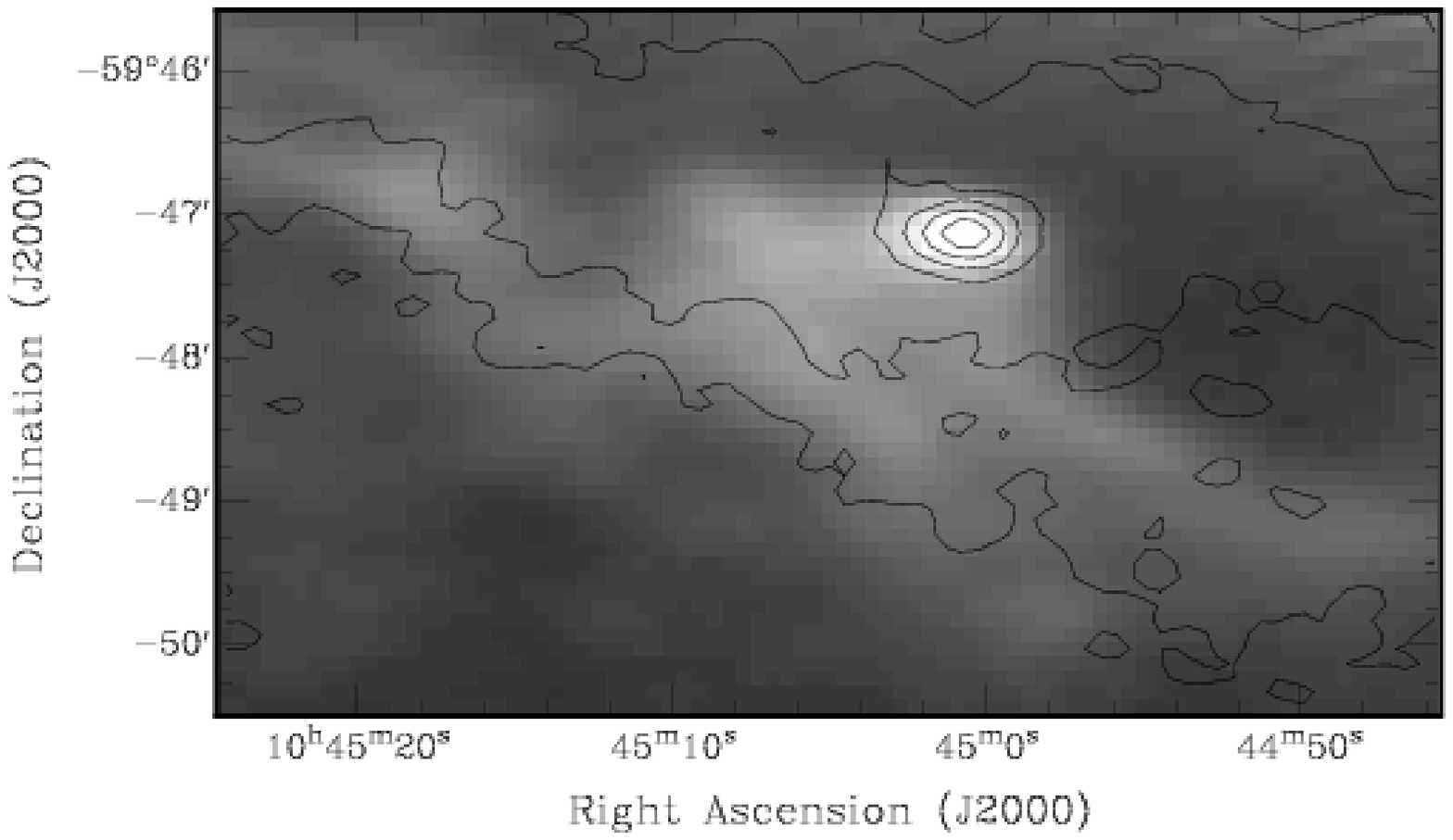,height=8.2cm,width=0.49\textwidth}}}
\caption{\label{south-region}The Southern Carina region. These images cover the same region as indicated in Fig.~\ref{pah-mosaic}.}
\end{figure*}

In all cases, the clumps show a strong coincidence between the 3.29- and 8-\um\, emission,
confirming the 8-\um\, emission detected here is dominated by PAH emission. Clump A2 is the 
only case where 21-\um\, emission is found above the background and associated with a known 
feature. The 21-\um\, emission seen toward this clump is faint and slightly offset from the 
peaks of the 3.29- and 8-\um\, emission and may not be associated directly with 
clump A2. Instead the feature may be associated with cooler dust belonging to clump 1
of \citeN{Cox951}. This molecular clump overlaps spatially with clump A2 and is distinguishable
via its more negative velocity (-34\kms). In the model by \citeN{Brooks00} clumps 1, 2 and 3 
(which correspond to the actual keyhole obscuring features seen at optical wavelengths) are
situated in front of the nebula. They show no signs of PDR emission and are therefore thought to
be less exposed to the radiation field coming from Tr 16.

\subsection{Southern region}
\label{southern-region}

Fig.~\ref{south-region}(a) shows 3.29-\um\, emission in grey scale with contours of 8-\um\, 
emission seen toward the southern region outlined in Fig.~\ref{pah-mosaic}. The 3.29-\um\, 
emission shows a strong source superposed on the extended emission associated with the edge of 
the large molecular cloud. The same is true for the 8-\um\, emission suggesting that the
emission is from PAHs. The brightest emission is coincident with the previously mentioned 
bright-rimmed molecular globule and in particular the source IRAS 10430-5931. 
This is a fairly faint \IRAS\, object with only a 12-\um\, upper limit in the \IRAS\, Point 
Point Source Catalog.  However, the IRAS Faint Source Catalog Reject file (FSCR)
gives S$_\nu$(12\um) = 6 Jy ($\pm$15\ per cent) and S$_\nu$(25\um) = 20 Jy ($\pm$11 
per cent). The corresponding 21-\um\, emission for this region is shown as contours in  
Fig.~\ref{south-region}(b). The source is a bright, compact source at 21-\um, making this the 
only site in the southern molecular cloud where there is a coincidence between an 
\IRAS\, source, 3.29- and 8-\um\, emission and a compact 21-\um\, source. 

\subsubsection{IRAS LRS spectrum for the southern globule}

\begin{figure}
\centering
\psfig{file=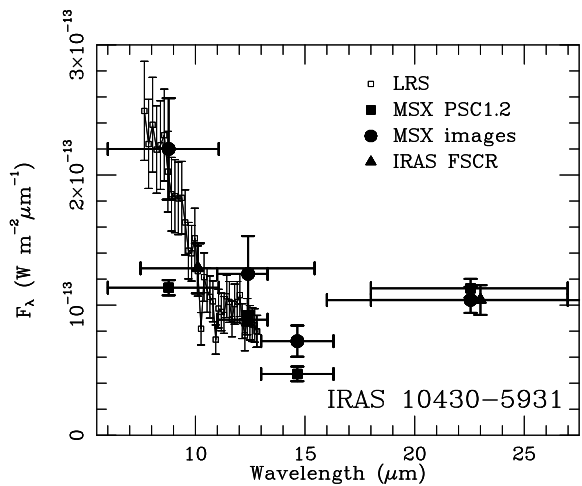,angle=-90,width=0.44\textwidth}
\caption{\label{lrs}
The IRAS LRS spectrum (from 7--13\um\, only) of IRAS 10430-5931 (open squares) with $\pm1\sigma$ 
error bars.  Plotted over the spectrum are the IRAS FSCR (filled triangles) and \MSX\, PSC1.2 
photometry points (filled squares), together with the results of spatial integrations of the 
source in the \MSX\, images (filled circles). All flux densities are plotted at the isophotal 
wavelengths appropriate to the source spectrum.}
\end{figure}

To investigate the character of the diffuse and point source emission further, 
we have sought \IRAS\, Low Resolution Spectrum (LRS) data. Three independent scans were 
extracted, averaged, recalibrated and normalised (as outlined in \citeNP{Cohen92}) to form 
the spectra shown in Fig.~\ref{lrs}. Unfortunately the long-wavelength part of the spectra
contain no valid data hence only the LRS data between 7 and 13\um\, are plotted here. 
Superposed on this spectrum are the flux densities  of IRAS 10430-5931 from the FSCR, the point 
source MSX5C\_G287.6393-00.7209 from the \MSX\, Point Source Catalogue (PSC1.2) and those that 
result from spatially integrating the corresponding diffuse peaks in the \MSX\, images. The 
inclusion of these data sets allows us to resolve the point sources (as seen by the smaller 
\MSX\, beam) from the extended emission (as seen by the larger beams of the FSCR and LRS). 

There are substantial differences between the point-source and the spatially 
integrated flux densities in the 8- and 12-\um\,bands. A modest difference occurs at 
14\um, with a good agreement between the pair of 21-\um\,flux densities. These confirm our 
suspicion that the diffuse contributions are chiefly due to emission from PAHs. On the larger 
spatial scale, the \IRAS\,12- and 25-\um\,flux densities agree well with those of the LRS 
spectrum, as do the \MSX\,diffuse measurements (the larger uncertainties with the latter are 
associated with the subtraction of the sky background).

The LRS appears to show a noisy PAH spectrum with a strong 7.7-\um\,band (defining the blue 
side of the spectrum), a marginal shoulder of emission near 8.7\um\,with no obvious fine 
structure emission lines. The steepness of the decline between 7.7 and 13\um\, rules out a 
stellar origin as it falls significantly faster than even the Rayleigh-Jeans domain in 
a cool stellar spectrum and lacks any photospheric absorptions.  No obvious 10-\um\, 
silicate absorption is seen, and the energy distributions of both the \MSX\, point
source and spatial integrations suggest thermal emission by cool dust grains.
The PAH feature at 11.3\um\, is not obvious in the LRS spectrum due to noisy data 
near 11\um.  However, its presence in the integrated diffuse \MSX\, emission 
is reasonably clear.  In \HII\, regions, one expects the intensity ratio of the 11.3-\um\, 
band to that of the powerful 7.7-\um\, band to be $\sim$0.29 (Cohen et al. 1986, 1989), but with
a range spanning over an order of magnitude among individual objects.  Similarly,
from the ratio of the 3.3- to 11.3-\um\, PAH features ($\sim2.2$: \citeN{Cohen86}), 
we can estimate the expected peak intensity of the 11.3-\um\, band.  Both these estimates 
indicate $\sim${\mbox{4 $\times$ $10^{-14}$ W m$^{-2}$ \um$^{-1}$}}, an amount that could well
be lost in the noise.

Therefore, we deduce that IRAS 10430-5931 has a compact component which
dominates the LRS spectrum but, blueward of $\sim$13\um, bright PAH emission bands   
selectively arise in the surrounding diffuse nebulosity.  The excellent agreement between the 
\IRAS\, 25-\um\, measurement and both \MSX\, point-source and spatially integrated data suggests 
that longward of $\sim$13\um, there is only a pointlike source.
We also note that the peak emission for all 4 \MSX\, bands in the IRAS 10430-5931 region occurs 
within one pixel of its neighbour, indicating a common point source (presumably the embedded 
star) with no evidence for any unusually spatially-extended emission at 21-\um. Without the 
long-wavelength LRS data we have no independent spectroscopy to offer but the accord of \MSX\, and 
\IRAS\, points near 23\um\, rule out the presence of a broad 22-\um\, emission line feature of the 
type found toward Car~I by \citeN{Chan00}.

\subsubsection{SED for the southern globule}
\label{southern-sed}

\begin{table*}
\begin{minipage}{170mm}
\begin{center}
\caption{\label{sed-properties}The derived parameters of the two-component black-body fits to the SEDs for the southern globule (IRAS 10430-5931) and selected sources seen in the northern region (N1, N2, N3 and N4) as shown in Figs.~\ref{south-sed} and~\ref{north-sed}. The parameters correspond to the temperature of the outer shell, \touter\,, at the maximum radius, \rmax,  and the temperature of the inner shell, \tinner\,, at the minimum radius, \rmin\,. Estimates of the luminosities (L$_{\mathrm TOT}$), spectral types and IR spectral indicies ($\alpha_{\mathrm IR}$) are also given.}
\begin{tabular}{@{}cccccccc}
\hline
Source  & \touter & \tinner & \rmin & \rmax & L$_{\mathrm TOT}$ & Spectral & $\alpha_{\mathrm IR}$ \\
        & (K) & (K)  & (au) & (au) & (10$^{4}$ \Lsun) & Type & \\
\hline	                      			      
IRAS 10430-5931 & 60 $\pm$ 4   & 430 $\pm$ 50 & 14 $\pm$ 3   & 4 760 $\pm$ 100 & 1.3$^b$ $\pm$ 0.6 & B0--0.5   & 3.8\\
N1 & 87 $\pm$ 10  & 330 $\pm$ 30 & 39  $\pm$ 6   & 1 510 $\pm$ 300 & 6.2 $\pm$ 0.6 & O8 	& 5.8\\
N2 & 74 $\pm$ 6   & 290 $\pm$ 20 & 55  $\pm$ 10  & 3 930 $\pm$ 200 & 20 $\pm$ 6  & O6--O6.5   & 7.1\\
N3 & 100 $\pm$ 15 & 2500$^a$    & 1.6 $\pm$ 0.8 & 640   $\pm$ 200 & 5.7 $\pm$ 0.7 & O8.5        &-1.7\\
N4 & 80 $\pm$ 10  & 380 $\pm$ 30 & 25 $\pm$ 10   & 1 440 $\pm$ 200 & 4.1  $\pm$ 0.5& O9--O9.5   &4.6\\
\hline
\end{tabular}
\end{center}
$^a$~lowest value for the temperature that fits the data given.\\
$^b$~this value is consistent with the lower limit estimate given by Megeath et al. (1996).
\end{minipage}
\end{table*}
 
\begin{figure}
\centering
\psfig{file=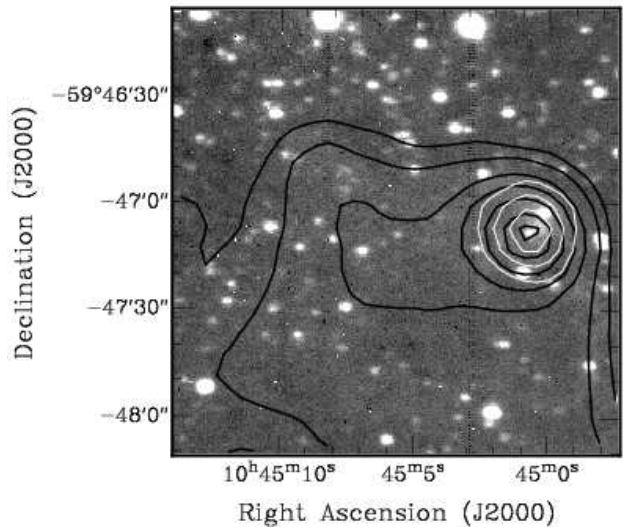,width=0.48\textwidth}
\caption{\label{south-k-AE}2.2-\um\, (K-band) image in grey scale of the southern globule. Black contours represent 8-\um\, emission; 1.2, 1.5, 2, 2.5, 3, 3.5, 3.8 $\times$ 10$^{-5}$ Wm$^{-2}$\,sr$^{-1}$. White contours represent 21-\um\, emission; 4, 5, 6 $\times$ 10$^{-5}$ Wm$^{-2}$\,sr$^{-1}$.}
\end{figure}

\citeN{Megeath96} found the southern bright-rimmed globule to have a velocity of $-$27\kms, a 
mass of 67\Msun and to be gravitationally bound. Also identified near the centre of the 
globule were four sources displaying anomalous near-IR colours, suggestive of deeply embedded 
pre-main sequence stars. To determine which of these corresponded to the 8- and
21-\um\, sources, our own K-band images were used. From Fig.~\ref{south-k-AE}
it is clear that the 8- and 21-\um\, emission peaks correspond to a strong K-band source.
This location corresponds to two of the four sources identified by \citeN{Megeath96} 
(sources 1 and 3). They are separated by less than 2 arcsec and are therefore unresolved in the
K-band image. The resolution of the \MSX\, images also makes differentiating between 
{\mbox{these sources difficult.}}

\begin{figure}
\centering
\psfig{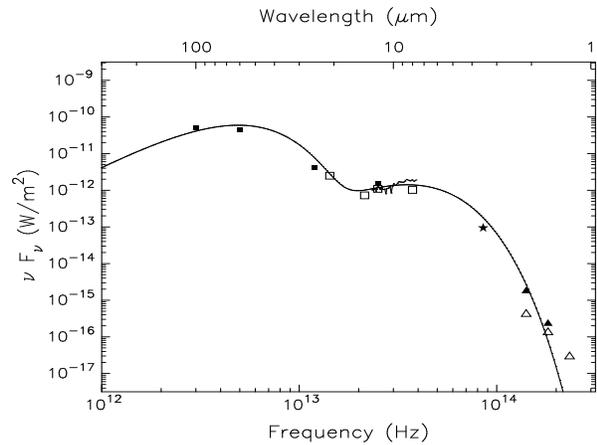}
\caption{\label{south-sed}Continuum SED for the southern globule. Filled squares represent \IRAS\, HIRAS data; open squares represent \MSX\, data; the star represents 3.5-\um\,data from SPIREX/Abu; and the filled and open triangles represent the near-IR data from Megeath's source 1 and 3 respectively. The thin line represents the \IRAS\, LRS spectrum for the globule, while the solid line represents the two-component black-body fit to the data (see Table~\ref{sed-properties}).} 
\end{figure}

Fig.~\ref{south-sed} shows the Spectral Energy Distribution (SED) for the globule. 
Included in this plot are fluxes from the \IRAS\, HIRAS catalogue for IRAS 10430-5931,
fluxes from the \MSX\, PSC, those calculated from the 3.5-\um\, SPIREX/Abu continuum data and 
those obtained from \citeN{Megeath96} for the sources 1 and 3. For the {\mbox {3.5-\um\,}} data, 
contamination from PAH emission at 3.29\um\, was estimated to be approximately 30 per cent 
and was removed from the continuum flux value before placing it on the SED. Also included 
on the SED is the \IRAS\, LRS for the source (thin line). The solid line is a two-component 
black-body fit to the data and represents the flux coming from two spherical shells of 
radius, \rmin\, and \rmax\, at corresponding temperatures \tinner\, and \touter\, respectively. 
This model is a simple representation of the emission seen toward an obscured source 
whose radiation is absorbed and re-emitted by a surrounding spherical dust shell. The 
temperature of the dust shell decreases radially outwards, the temperatures \tinner\,
and \touter\, can be determined, along with the ratio of the angular sizes of the 
dust shells ((\rmin/\rmax)$^{2}$) from this model. Using the known distance to the source,
the luminosity and the derived temperatures, \rmin\, and \rmax\, can be determined. The 
parameters obtained from fits to this SED are given in Table~\ref{sed-properties}. 

Emission from PAH and absorption from silicate molecules is important to consider for
this model. Contamination from the former was removed from the data, but the latter remains in 
the SED. This is clearly seen as a dip in the spectrum, which is significant when attempting 
to determine the best fit. Although this is a simple fit to the temperatures and radii, this 
model allows us to obtain a first order estimation to the conditions within these regions.

\begin{figure*}
\centering
\mbox{\subfigure[3.29-\um\,emission in grey scale with contours of 8-\um\, emission. Levels are 2.7, 3.5, 4.5, 5.4, 6.5, 7.3, 8.1, 8.9, 9.7 $\times$ 10$^{-5}$ Wm$^{-2}$\,sr$^{-1}$.]{\psfig{file=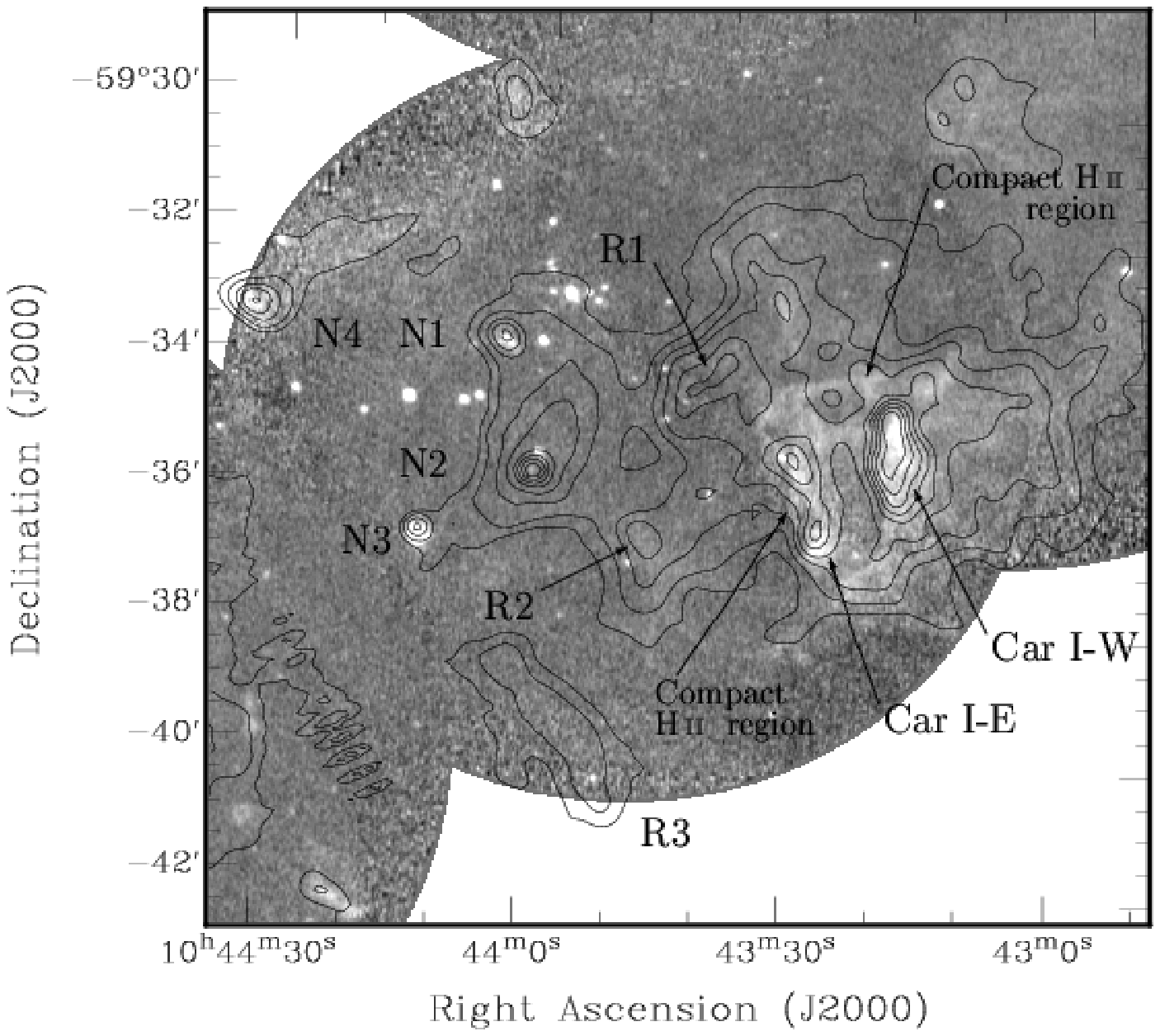,width=0.48\textwidth}}
\subfigure[8-\um\,emission in grey scale with contours of 21-\um\, emission. Levels are 0.6, 1.1, 1.6, 2.1, 2.6, 3.1, 3.6 $\times$ 10$^{-4}$ Wm$^{-2}$\,sr$^{-1}$.]{\psfig{file=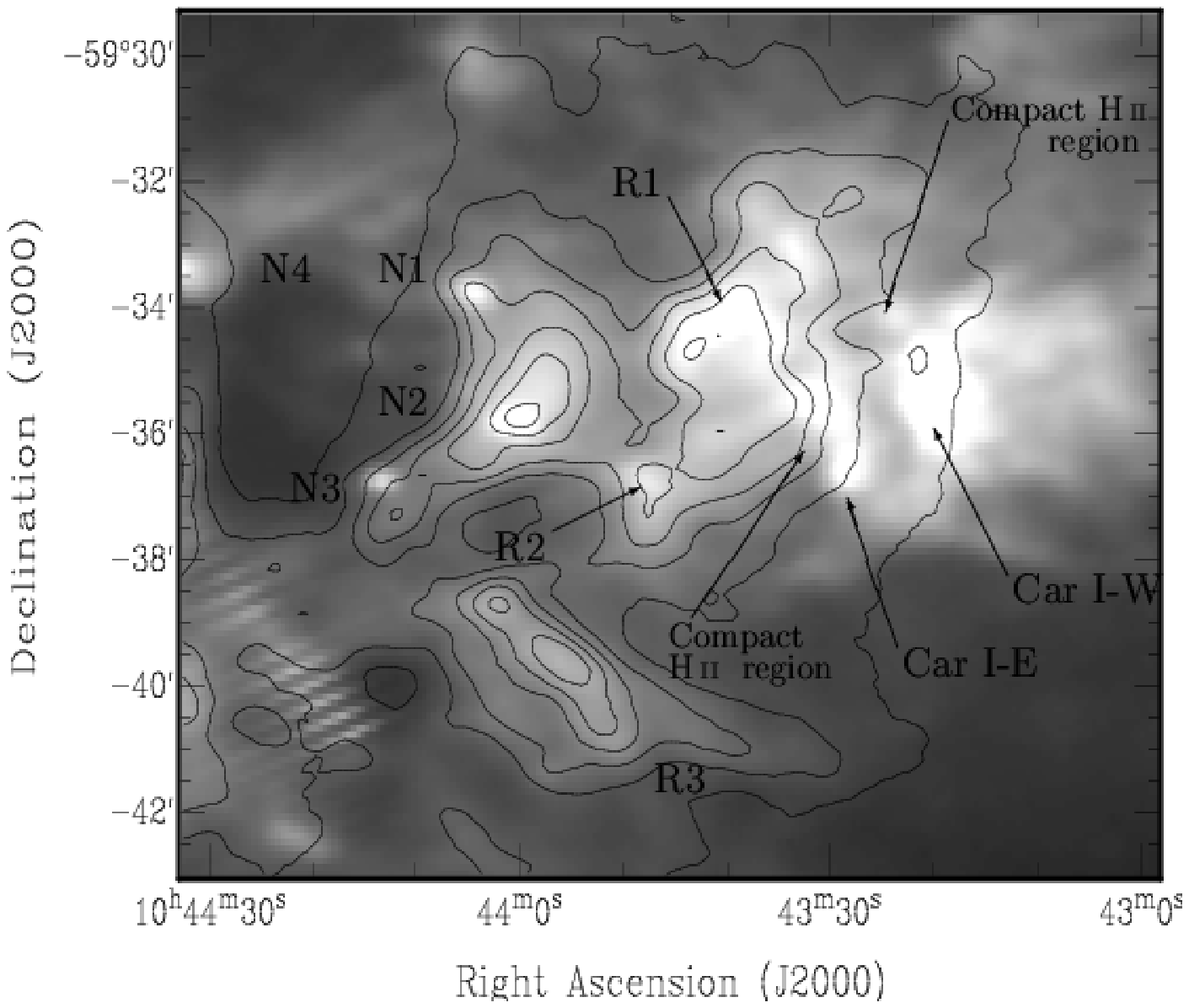,width=0.5\textwidth}}}
\caption{\label{north-region}The Northern Carina region. Several point sources are labelled on the image (N1, N2, N3 and N4), along with regions of diffuse emission (R1, R2, and R3), as well as the ionization fronts and compact \HII\, regions identified by Brooks \& Whiteoak (2001). The parallel stripes seen in the contours of (a) and the grey scale image of (b) are artifacts from diffractions of \nCar. These images cover the same region as indicated in Fig.~\ref{pah-mosaic}.}
\end{figure*}

The shape of the SED is consistent with a Class ${\mathrm\scriptstyle I}$ low-mass
YSO in the classification scheme of Lada (1987) (these classifications are only used in the 
current study to describe the shape of the SEDs as it is not clear if high-mass stars follow a 
similar evolutionary process). The shape of the SED is determined from the
spectral index ($\alpha_{\mathrm IR}$), which is a measure of the slope of the SED between 
near- and mid-IR wavelengths (in this case between 2 and 21-\um). Although the SED rises in 
the far-IR, the exact region where the emission peaks is unclear from these data alone. It may 
be that the emission continues to increase into the sub-millimetre regime, thus corresponding 
to a lower outer temperature and higher total luminosity.

\citeN{Megeath96} also constructed a SED for this source, including only the \IRAS\, and near-IR
data points. From this limited information they put forward one possible explanation for the
shape of the SED suggesting that the globule contained an ultra-compact (UC) \HII\, region. From 
the additional data shown here it appears this is indeed the case. The shape of the SED and 
the temperatures, radii and luminosities derived from it are all consistent with other 
UC \HII\,regions identified (e.g. \citeNP{Walsh99}). The \MSX\, colours (8$-$12\um), 
(8$-$14\um) and (8$-$21\um) of 1.39\,mag, 1.50\,mag and 4.08\,mag are also consistent with 
those properties of UC \HII\, regions as determined from infrared models of the point-source sky
(\citeNP{Wainscoat92}; \citeNP{Cohen93}). Radio continuum images of the region however
show no source coincident with the globule \cite{Whiteoak94}. The luminosity derived from 
the SED corresponds to a star of spectral type B0.5 \cite{Panagia73}. This most likely 
corresponds to Megeath's source 1 (filled triangles in Fig.~\ref{south-sed}) as this source 
appears to be more consistent with the overall fit. In order to make a more confident 
distinction between the two sources however, higher resolution imaging is required.

\subsection{Northern region}
\label{northern-region}

As noted earlier, the 3.29-\um\, emission in the northern part of the Carina Nebula extends 
over a large part of the northern molecular cloud and is brightest near the \HII\, region Car~I.
Fig.~\ref{north-region}(a) shows the 3.29- and 8-\um\, emission of this region in detail.
The brightest 3.29-\um\, emission is concentrated at the edge of the molecular cloud, with 
faint diffuse emission seen extending across the region. In contrast, the {\mbox {8-\um\,}} 
emission is much more diffuse and contains strong emission over the whole region, including those 
areas where there is faint 3.29-\um\, emission. Fig.~\ref{north-region}(b) provides a
comparison of the 8- and 21-\um\, emission. The emission structures coincide very closely, 
with the exception of the region containing strong 3.29-\um\, emission. Here there is almost
no 21-\um\, emission and the bright 3.29-\um\, emission can be largely attributed to the PAH
emission line. The regions labelled R1, R2 and R3 are the contrary. These regions show
little 3.29-\um\, emission, with a strong coincidence between the 8- and 21-\um\, emission 
structures. This suggests the emission is from dust at these locations and most likely represents 
the back face of the molecular cloud. 

\subsubsection{The Car~I \HII\, region}

\begin{figure*}
\centering
\psfig{file=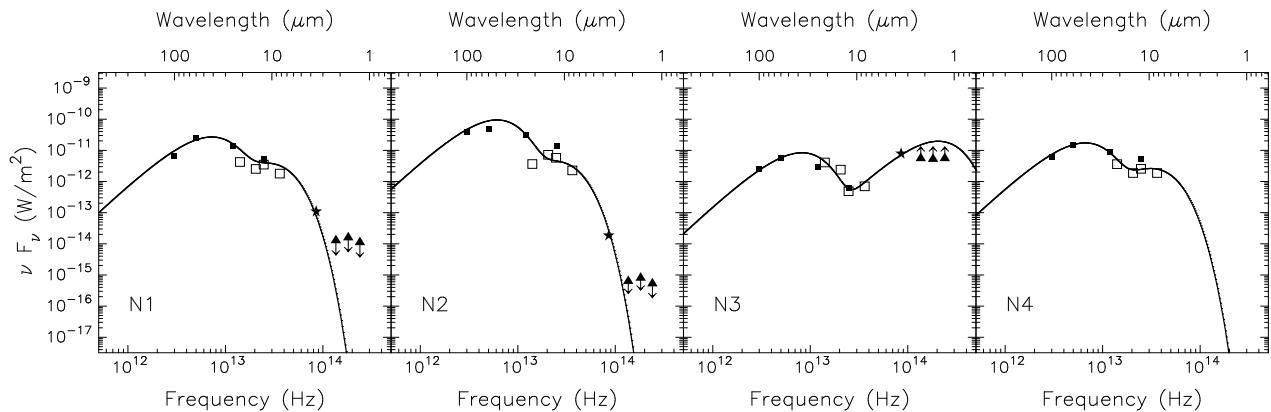,angle=-90,width=\textwidth}
\caption{\label{north-sed}SEDs for sources seen in the northern region. Filled squares represent \IRAS\, data; open squares represent \MSX\, data; the star represents 3.5-\um\, data from SPIREX/Abu; and filled triangles represent near-IR data from CASPIR. The solid line is the resulting two-component black-body fits to the data (see Table~\ref{sed-properties}).}
\end{figure*}

Car~I contains several ionization fronts arising from Tr 14 the most prominent being Car~I-E
and Car~I-W (\citeNP{Whiteoak94}; Brooks et al. 2001). Car~I-E coincides with an optically 
bright-rimmed edge of the molecular cloud. Bright 3.29-\um\, emission also traces this edge
and forms a series of linear ridges that point toward Tr 14. One of the southern ridges 
corresponds to a strong region of 8-\um\, emission (labelled as Car~I-E in 
Fig.~\ref{north-region}). Car~I-W is situated further west and curves around a dense 
molecular clump. Bright 3.29- and 8-\um\, emission also trace this ionization front (labelled 
as Car~I-W in Fig.~\ref{north-region}). The absence of a strong coincidence between the 
8- and 21-\um\, emission in both Car~I-E and Car~I-W confirms that the 8-\um\, emission can be 
attributed to PAH molecules. These results are consistent with a model proposed by 
Brooks et al. (2001) stating that Car~I-E is interacting with the front face of the GMC and 
creating a PDR seen edge-on, while Car~I-W is interacting with the back face and 
creating a second PDR seen face on.
%{\mbox{second PDR seen face on.}}

\subsubsection{Point Sources and their SEDs}
The two compact \HII\, regions identified by Brooks et al. (2001) are located at the edges 
of the ionization fronts Car~I-E and Car~I-W (labelled in Fig.~\ref{north-region}). The 
8-\um\, emission associated with Car~I-E actually curves around the compact source located here. 
The second compact \HII\, region associated with the edge of Car~I-W further to the north, 
shows a coincidence with 3.29-\um\, emission, faint 8-\um\, emission but very little 
21-\um\, emission. 

Several new point sources are evident in the 8-\um\,image (labelled N1, N2, N3 and N4) and
are located on the edges of a small region completely devoid of all IR emission. The 
sources N1 and N2 correspond to extended sources in the 3.29-\um\, image, with differing 
amounts of heated dust extending around them (as traced by the 21-\um\, emission). The 
source N3 corresponds to a strong point source in {\mbox {3.29-\um\,}} emission, with little 
21-\um\, emission. A peak in 21-\um\, emission exists slightly offset from this position, 
however it is not clear if this feature is related to the 8-\um\, source. The source N4 shows 
a strong coincidence between the 3.29- and 8-\um\, structures, with little 21-\um\, emission 
associated with it, suggesting the emission is from PAH molecules. 

The spatial offsets between the 8- and 21-\um\, peaks of these sources might be attributable to 
contamination of the 21-\um\,  contours by a strong 22-\um\,  dust emission feature. 
If this feature exists across the whole Car~I region and is excited by either Tr 14 or \nCar, 
then the amount of contamination determined from the Chan \& Onaka spectra will only serve as a 
lower limit (as estimated in section~\ref{msx-obs}). The point sources identified here are 
located closer to both possible sources than the positions at which the spectra were taken and 
thus would be more likely to have a higher amount of the 22-\um\, dust feature underlying the 
overall emission. Without spectra nearer to the individual point sources, it is difficult to 
estimate the contamination seen in the 21-\um\, band due to the underlying 22-\um\, dust emission 
feature.

The SEDs for the point sources N1, N2, N3 and N4 are shown in Fig.~\ref{north-sed}. 
Due to the complexity and confusion in this region, several of these sources contained
flags in the \MSX\, PSC1.2 indicating the emission was a lower limit; a result of their extended
nature. For these cases, the flux over the region approximating the 8-\um\,
peak was estimated and a nearby background removed. To confirm that this method derived
acceptable flux estimates, the flux was also measured for the sources in the bands where 
there was data in the PSC1.2. The estimation of the flux was, in the worst case, a factor of 2 
larger than the correct value. This method was also used to estimate the \IRAS\, fluxes at
these locations. Several \IRAS\, sources exist in this region, but none were found to be
consistent with the 8-\um\, point sources. The effects of contamination from 3.29-\um\, PAH 
emission for the continuum fluxes obtained at 3.5\um\, were also estimated for these sources 
(30 per cent for N1, 5 per cent for N2 and 1 per cent for N3) and were removed from the flux 
value before placing the point on the SED. The coincidence between the \MSX\,  and \IRAS\, points 
near 21-\um\, for the sources N3 and N4 suggest the possible contamination from the 22-\um\, dust 
emission feature is small. However, for the source N1 and N2, the differing flux values could 
possibly result from a strong 22-\um\, dust emission feature seen near these sources. A 
quantitative estimate of such contamination however is difficult with the current data set.

Near-IR images of the sources N1 and N2 reveal extended diffuse emission. The fluxes
shown in the SEDs for these wavebands are representations of the emission at these 
points and are upper limits to any source that may be associated with the 8-\um\, peak.
For N3 there exists a strong source in the near-IR images, saturating in our images, hence
the flux plotted on the SED for this source represents a lower limit to the actual flux. The 
source N4 was outside our observed regions, but shows a very red and extended region in 
2MASS\footnote{For more information about the 2MASS project and database, see http://www.ipac.caltech.edu/2mass/index.html} 
three colour images of the region, consistent with the SED fit at these wavelengths.

The solid lines in these plots represent the two-component black-body fits as described in 
section~\ref{southern-sed}. The parameters from these simple fits are given in 
Table~\ref{sed-properties}. The sources N1, N2 and N4 have similar properties to 
the SED for the southern globule; they peak in the far-IR, correspond to similar temperatures
and radii, have $\alpha_{\mathrm IR} >$ 0 and luminosities that correspond to early type stars 
(in this case O6--O9.5). Most likely these sources correspond to UC \HII\, regions.
Source N3 however displays the same features in the far-IR as the other sources but 
shows remarkably different characteristics in the thermal- and near-IR regimes. This results 
in an $\alpha_{\mathrm IR}$ of -1.7, similar in shape to a 
{\mbox{Class ${\mathrm\scriptstyle II}$\,}} SED and thus more evolved than the other sources 
in the region. These objects are thought to represent young stars
surrounded by a circumstellar disc of material \cite{Lada87}. The fact that 
there is emission in the far-IR suggests that cool dust still exists around this source and is
therefore consistent with this picture.

\section{Conclusions}
We have obtained high-resolution, wide-field 3.29-\um\, emission images across the Carina
Nebula in order to trace the PDRs thought to be widespread in regions of massive star 
formation. Combined with \MSX\, 8--21\um\, data, CASPIR 2.2-\um\, data and SEST 
molecular line data, these images emphasize the three different environments of the Carina Nebula: 
the Keyhole, southern and northern molecular cloud regions.

\subsection{Keyhole region}
The Keyhole region contains many of the discrete clumps identified in our images, several of 
which have been studied previously. Two additional clumps were identified via their 3.29-\um\,
PAH emission and mapped in \coh\, and \col. Combined with published data for the other clumps 
in the region, it is clear that all the molecular material in this region is exposed to the 
same intense FUV radiation. The high \coh\, and (1--0) line ratios, as well as significantly 
higher Virial masses compared to the LTE mass estimates, suggests that all the clumps are 
externally heated with PDRs on their surfaces and supported by external pressure. This is 
consistent with suggestions that the clumps are photo-evaporating and represent the swept-up 
remains of the molecular material from which \nCar\, formed. The clumps also follow the 
geometrical model developed by \citeN{Brooks00} relating their optical appearance, PAH 
emission structures and velocities to their location within the region and with respect to 
\nCar. While these new clumps are consistent with the above-mentioned properties, there are 
certain differences that set them apart. Both of these structures have higher PAH fluxes and 
larger sizes relative to the other clumps in the region. This suggests that the 
photo-evaporation process may be slower for these clumps, a direct consequence of their
further displacement from \nCar.

\subsection{Southern region}
Toward the southern region the images reveal an externally heated globule with
a correspondence between strong 3.29-, 8- and 21-\um\, emission. The morphology of the globule 
suggests that the winds and FUV radiation from \nCar\, and Tr 16 may be responsible for 
carving the molecular material around a dense compact region, penetrating into the molecular 
cloud and forming a PDR on its surface. The location is also 
consistent with IRAS 10430-5931 and four sources displaying anomalous near-IR colours. 
By comparison with the locations of the near-IR sources and the mid-IR data, we were able
to eliminate two of these sources as producing the observed fluxes. The SED shows a peak in 
the far-IR, falling off at shorter wavelengths. Through the use 
of a two-component black-body fit to this SED we have tentatively identified the 
emission as coming from a single source, most likely corresponding to an UC \HII\, region.

\subsection{Northern region}
The northern region displays more varied phenomena than either the Keyhole or southern regions.
Here, we see regions of strong PAH emission and regions containing heated dust, both 
separately and inter-mixed. The diffuse PAH emission we see is consistent with the edge 
of the molecular cloud and with the strong ionization fronts Car~I-E and Car~I-W. The 
coincidence between the 3.29- and 8-\um\, data at these locations confirms the existence 
of PDRs at these interfaces.  

Several strong 8-\um\, point sources are evident in this region and are located on the edge 
of a small region completely devoid of extended 3.29-, 8- and 21-\um\, emission. SEDs for these 
sources and two-component black-body model fits to these data suggest three of the four sources 
are Class ${\mathrm \scriptstyle I}$ protostars. The fourth source has a different SED,
peaking in the near-IR/optical regime with significant far-IR emission. This SED is
characteristic of a more evolved Class ${\mathrm \scriptstyle II}$ protostar, containing a 
young star surrounded by a circumstellar disc (rather than completely embedded).

\subsection{Ongoing star formation in the Carina Nebula?}
From the data presented here it is clear that star formation within the Carina Nebula has not
been completely halted by the clusters of young massive stars. This supports the changing view 
that the Carina Nebula contains active star formation, contrary to previous conclusions.
The northern region was found to contain more sites of massive star formation compared to the 
southern and Keyhole regions. All young stars identified are associated with UC \HII\, regions and 
are possibly a result of triggered star formation. The four sources identified in the northern 
region for instance, all lie on the edges of a small cavity devoid of IR emission.

The differing environments of these three regions contributes significantly to the star formation 
activity seen within them. The Keyhole region shows the effects of such a close proximity 
to \nCar, the environment being too harsh for any star formation activity. It may be however, 
that the massive stellar members of Tr 16 including \nCar, are contributing to star formation at 
the edge of the southern molecular cloud. It appears as if the winds and radiation from these 
massive stars may be disrupting the molecular cloud and inducing star formation within a globule 
on the edge. The environment in the northern region produced by Tr 14 is also extremely 
disruptive on the surrounding GMC. The location and identification of the star formation 
activity, PDRs and greater amounts of heated dust in this region is consistent with a younger 
age for Tr 14 compared to Tr 16 and the southern molecular cloud. Thus the environment in the 
Keyhole and southern regions may represent a future state for the northern region.

\section*{Acknowledgments}
We would like to thank Andrew Walsh and Angie Schultz for invaluable discussions, colleagues 
from CARA for their support in operating the facilities at the South Pole, in particular
Charley Kaminski for taking the SPIREX/Abu data. We acknowledge the support of the Australian
Research Council and travel funding obtained through ANSTO. MC thanks NASA for supporting 
this work through its Long Term Space Astrophysics programme, under grant NAG5-7936 with UC 
Berkeley.

%\bibliography{mnrasmnemonic,references}

\begin{thebibliography}{}

\bibitem[\protect\citeauthoryear{Allamandola, Tielens, \& Barker}{Allamandola
  et~al.}{1985}]{Allamandola85}
Allamandola L.~J., Tielens A.~G. G.~M.,  Barker J.~R., 1985, ApJ, 290, L25

\bibitem[\protect\citeauthoryear{Allamandola, Tielens, \& Barker}{Allamandola
  et~al.}{1989}]{Allamandola89}
Allamandola L.~J., Tielens A.~G. G.~M.,  Barker J.~R., 1989, ApJS, 71, 733

\bibitem[\protect\citeauthoryear{Arendt, Dwek, \& Moseley}{Arendt
  et~al.}{1999}]{Arendt99}
Arendt R., Dwek E.,  Moseley S., 1999, ApJ, 521, 234

\bibitem[\protect\citeauthoryear{{Brooks}}{{Brooks}}{2000}]{Brooks-phd}
{Brooks} K.~J., 2000, Ph.D. thesis, University of New South Wales, Australia

\bibitem[\protect\citeauthoryear{Brooks et~al.}{Brooks et~al.}{2000}]{Brooks00}
Brooks K.~J., Burton M.~G., Rathborne J.~M., Ashley M.~C.~B.,  Storey J.~W.
  V.~S., 2000, MNRAS, 319, 95

\bibitem[\protect\citeauthoryear{Brooks, Storey \& Whiteoak}{Brooks
  et~al.}{2001}]{Brooks01}
Brooks K.~J., Storey J.~W.~V.,  Whiteoak J.~B., 2001, MNRAS, in press

\bibitem[\protect\citeauthoryear{Brooks, Whiteoak \& Storey}{Brooks
  et~al.}{1998}]{Brooks98}
Brooks K.~J., Whiteoak J.~B.,  Storey J.~W.~V., 1998, Proc.\ Astron.\ Soc.\
  Aust., 15(2), 202

\bibitem[\protect\citeauthoryear{Buisson et~al.}{Buisson
  et~al.}{1999}]{Buisson99}
Buisson G., Desbats L., Duvert G., Forveille T., Gras R., Guilloteau S., Luca
  R.,  Valiron P., 1999, GILDAS -- An introduction to Grenoble and Line Data
  Analysis System, see http://iram.fr/GS/gildas.html

\bibitem[\protect\citeauthoryear{Burton et~al.}{Burton et~al.}{2000}]{Burton00}
Burton M.~G. et~al., 2000, ApJ, 542, 359

\bibitem[\protect\citeauthoryear{Castets et~al.}{Castets
  et~al.}{1990}]{Castets90}
Castets A., Duvert G., Dutrey A., Bally J., Langer W.~D.,  Wilson R.~W., 1990,
  A\&A, 234, 469

\bibitem[\protect\citeauthoryear{Chan \& Onaka}{Chan \& Onaka}{2000}]{Chan00}
Chan K.~W.,  Onaka T., 2000, ApJ, 533, L33

\bibitem[\protect\citeauthoryear{Cohen}{Cohen}{1993}]{Cohen93}
Cohen M., 1993, AJ, 105, 2860

\bibitem[\protect\citeauthoryear{Cohen et~al.}{Cohen et~al.}{1986}]{Cohen86}
Cohen M., Allamandola L.~J., Tielens A.~G. G.~M., Bregman J., Simpson J.~P.,
  Witteborn F.~C., Wooden D.,  Rank D.~M., 1986, ApJ, 302, 737

\bibitem[\protect\citeauthoryear{Cohen et~al.}{Cohen et~al.}{1989}]{Cohen89}
Cohen M., Tielens A.~G. G.~M., Bregman J., Witteborn F.~C., Rank D.~M.,
  Allamandola L.~J., Wooden D.,  de~Muizon M., 1989, ApJ, 341, 246

\bibitem[\protect\citeauthoryear{Cohen, Walker, \& Witteborn}{Cohen
  et~al.}{1992}]{Cohen92}
Cohen M., Walker R.~G.,  Witteborn F.~C., 1992, AJ, 104, 2030

\bibitem[\protect\citeauthoryear{Cox}{Cox}{1995}]{Cox952}
Cox P., 1995, RevMexAA (Serie de Conferencias), 2, 105

\bibitem[\protect\citeauthoryear{Cox \& Bronfman}{Cox \&
  Bronfman}{1995}]{Cox951}
Cox P.,  Bronfman L., 1995, A\&A, 299, 583

\bibitem[\protect\citeauthoryear{de~Graauw et~al.}{de~Graauw
  et~al.}{1981}]{deGraauw81}
de~Graauw T., Lidholm S., Fitton B., Beckman J., Irael F.~P., Nieuwenhuijzen
  H.,  Vermue J., 1981, A\&A, 102, 257

\bibitem[\protect\citeauthoryear{Egan et~al.}{Egan et~al.}{1998}]{Egan98}
Egan M.~P., Shipman R.~F., Price S.~D., Carey S.~J., Clark F.~O.,  Cohen M.,
  1998, ApJ, 494, L199

\bibitem[\protect\citeauthoryear{Fowler et~al.}{Fowler et~al.}{1998}]{Fowler98}
Fowler A.~M. et~al., 1998, Proc. SPIE, 3354, 1170

\bibitem[\protect\citeauthoryear{Gardner \& Morimoto}{Gardner \&
  Morimoto}{1968}]{Gardner68}
Gardner F.~F.,  Morimoto M., 1968, ApJ, 21, 881

\bibitem[\protect\citeauthoryear{Geballe et~al.}{Geballe
  et~al.}{1994}]{Geballe94}
Geballe T.~R., Joblin C., d'Hendecourt L., Jourdain~de Muizon M., Tielens A.~G.
  G.~M.,  L\'{e}ger A., 1994, ApJ, 434, L15

\bibitem[\protect\citeauthoryear{Genzel et~al.}{Genzel et~al.}{1988}]{Genzel88}
Genzel R., Harris A.~I., Stutzki J.,  Jaffe D.~T., 1988, ApJ, 332, 1049

\bibitem[\protect\citeauthoryear{Gooch}{Gooch}{1996}]{Gooch96}
Gooch R., 1996, in Jacoby G.~H.,  Barnes J., ed, Astronomical Data Analysis
  Software and Systems V, Vol. 101.
\newblock ASP Conference Series, San Francisco, p.~80

\bibitem[\protect\citeauthoryear{Grabelsky et~al.}{Grabelsky
  et~al.}{1988}]{Grabelsky88}
Grabelsky D.~A., Cohen R.~S., Bronfman L.,  Thaddeus P., 1988, ApJ, 331, 181

\bibitem[\protect\citeauthoryear{Hasegawa}{Hasegawa}{1996}]{Hasegawa96}
Hasegawa T., 1996, in Latter W.~B., Radford S.~J.~E., Jewell P.~R., Mangum
  J.~G.,  Bally J., ed, CO: Twenty-five Years of Millimeter-Wave Spectroscopy -
  Proceedings of the 170th Symposium of the IAU.
\newblock Kluwer, Dordrecht, p.~39

\bibitem[\protect\citeauthoryear{Hereld et~al.}{Hereld et~al.}{1990}]{Hereld90}
Hereld M., Rauscher B.~J., Harper D.~A.,  Pernic R.~J., 1990, Proc. SPIE, 1235,
  43

\bibitem[\protect\citeauthoryear{Hollenbach \& Tielens}{Hollenbach \&
  Tielens}{1999}]{Hollenbach99}
Hollenbach D.,  Tielens A.~G. G.~M., 1999, Rev Mod Phys, 71, 173

\bibitem[\protect\citeauthoryear{Joblin et~al.}{Joblin et~al.}{1995}]{Joblin95}
Joblin C., Tielens A.~G. G.~M., Allamandola L.~J., L\'{e}ger A., d'Hendecourt
  L., Geballe T.~R.,  Boissel P., 1995, Planet.\ Space Sci., 43, 1189

\bibitem[\protect\citeauthoryear{Jones et~al.}{Jones et~al.}{1999}]{Jones99}
Jones A., Frey V., Verstraete L., Cox P.,  Demyk K., 1999, in Cox P.,  Kessler
  M., ed, The Universe as Seen by ISO.
\newblock ESA SP-427; Noordwijk: ESA, p. 679

\bibitem[\protect\citeauthoryear{Lada}{Lada}{1987}]{Lada87}
Lada C., 1987, in Peimbert M.,  Jugaku J., ed, IAU Symp. 115, Star Forming
  Regions.
\newblock Kluwer, Dordrecht, p.~1

\bibitem[\protect\citeauthoryear{L\'{e}ger \& Puget}{L\'{e}ger \&
  Puget}{1984}]{Leger84}
L\'{e}ger A.,  Puget J.~L., 1984, A\&A, 137, L5

\bibitem[\protect\citeauthoryear{McGregor et~al.}{McGregor
  et~al.}{1994}]{McGregor94}
McGregor P., Hart J., Downing M., Hoadley D.,  Bloxham G., 1994, Experimental
  Astronomy, 3, 139

\bibitem[\protect\citeauthoryear{McGregor}{McGregor}{1995}]{McGregor95}
McGregor P.~J., 1995, CASPIR data reduction routines, see
  http://msowww.anu.edu.au/observing/docs/manual/manual.html

\bibitem[\protect\citeauthoryear{Megeath et~al.}{Megeath
  et~al.}{1996}]{Megeath96}
Megeath S.~T., Cox P., Bronfman L.,  Roelfsema P.~R., 1996, A\&A, 305, 296

\bibitem[\protect\citeauthoryear{Meixner et~al.}{Meixner
  et~al.}{1992}]{Meixner92}
Meixner M., Haas M.~R., Tielens A.~G. G.~M., Erickson E.~F.,  Werner M., 1992,
  ApJ, 390, 499

\bibitem[\protect\citeauthoryear{Mill et~al.}{Mill et~al.}{1994}]{Mill94}
Mill J.~D., O'Neil R.~R., Price S., Romick G.~J., Uy~O.~M.,  Gaposchkin E.~M.,
  1994, J. Spacecraft Rockets, 31, 900

\bibitem[\protect\citeauthoryear{Moutou et~al.}{Moutou et~al.}{2000}]{Moutou00}
Moutou C., Verstraete L., L{\'e}ger A., Sellgren K.,  Schmidt W., 2000, A\&A,
  354, L17

\bibitem[\protect\citeauthoryear{Panagia}{Panagia}{1973}]{Panagia73}
Panagia N., 1973, ApJ, 78, 929

\bibitem[\protect\citeauthoryear{Price et~al.}{Price et~al.}{2001}]{Price01}
Price S.~D., Egan M.~P., Carey S.~J., Mizuno D.,  Kuchar T., 2001, AJ, 121,
  2819

\bibitem[\protect\citeauthoryear{Retallack}{Retallack}{1983}]{Retallack83}
Retallack D.~S., 1983, MNRAS, 204, 669

\bibitem[\protect\citeauthoryear{Sakamoto et~al.}{Sakamoto
  et~al.}{1994}]{Sakamoto94}
Sakamoto S., Hayashi M., Hasegawa T., Handa T.,  Oka T., 1994, ApJ, 425, 641

\bibitem[\protect\citeauthoryear{Schneider et~al.}{Schneider
  et~al.}{1998}]{Schneider98}
Schneider N., Stutzki J., Winnewisser G., Poglitsch A.,  Madden S., 1998, A\&A,
  338, 262

\bibitem[\protect\citeauthoryear{Scoville et~al.}{Scoville
  et~al.}{1991}]{Scoville91}
Scoville N.~Z., Sargent A.~I., Sanders D.~B.,  Soifer B.~T., 1991, ApJ, 366, L5

\bibitem[\protect\citeauthoryear{Sellgren, Werner, \& Dinerstein}{Sellgren
  et~al.}{1983}]{Sellgren83}
Sellgren K., Werner M.~W.,  Dinerstein H.~L., 1983, ApJ, 217, L13

\bibitem[\protect\citeauthoryear{Smith et~al.}{Smith et~al.}{2000}]{Smith00}
Smith N., Egan M.~P., Carey S., Price S.~D., Morse J.~A.,  Price P.~A., 2000,
  ApJ, 532, L145

\bibitem[\protect\citeauthoryear{Stutzki et~al.}{Stutzki
  et~al.}{1988}]{Stutzki88}
Stutzki J., Stacey G.~J., Genzel R., Harris A.~I., Jaffe D.~T.,  Lugten J.~B.,
  1988, ApJ, 332, 379

\bibitem[\protect\citeauthoryear{Tauber et~al.}{Tauber et~al.}{1995}]{Tauber95}
Tauber J.~A., Lis D.~C., Keene J., Schilke P.,  Buettgenbach T.~H., 1995, A\&A,
  297, 567

\bibitem[\protect\citeauthoryear{Tovmassian}{Tovmassian}{1995}]{Tovmassian95}
Tovmassian H.~M., 1995, RevMexAA (Serie de Conferencias), 2, 83

\bibitem[\protect\citeauthoryear{Ulich \& Haas}{Ulich \& Haas}{1976}]{Ulich76}
Ulich B.~L.,  Haas R.~W., 1976, ApJ, 30, 247

\bibitem[\protect\citeauthoryear{Verstraete et~al.}{Verstraete
  et~al.}{2001}]{Verstraete01}
Verstraete L. et~al., 2001, A\&A, 372, 981

\bibitem[\protect\citeauthoryear{Wainscoat et~al.}{Wainscoat
  et~al.}{1992}]{Wainscoat92}
Wainscoat R.~J., Cohen M., Volk K., Walker H.~J.,  Schwartz D.~E., 1992, ApJS,
  83, 111

\bibitem[\protect\citeauthoryear{Walborn}{Walborn}{1995}]{Walborn95}
Walborn N.~R., 1995, RevMexAA (Serie de Conferencias), 2, 51

\bibitem[\protect\citeauthoryear{Walsh et~al.}{Walsh et~al.}{1999}]{Walsh99}
Walsh A.~J., Burton M.~G., Hyland A.~R.,  Robinson G., 1999, MNRAS, 309, 905

\bibitem[\protect\citeauthoryear{Whiteoak \& Otrupcek}{Whiteoak \&
  Otrupcek}{1984}]{Whiteoak84}
Whiteoak J.~B.,  Otrupcek R.~E., 1984, Proc.\ Astron.\ Soc.\ Aust., 5(4), 552

\bibitem[\protect\citeauthoryear{Whiteoak}{Whiteoak}{1994}]{Whiteoak94}
Whiteoak J.~B.~Z., 1994, ApJ, 429, 225

\end{thebibliography}
%\bibliographystyle{mnras}

\label{lastpage}
\end{document}